\documentclass[
aps,
pra,
superscriptaddress,
showpacs,
showkeys,
notitlepage,
twocolumn, 
numbers,
longbibliography,
nofootinbib]
{revtex4-1}

\usepackage[utf8]{inputenc}
\usepackage{babel}
\usepackage{mathtools}
\usepackage{amsmath}
\usepackage{caption}
\usepackage{subcaption}
\usepackage{amssymb}
\usepackage{graphicx}
\usepackage{tikz}
\usepackage{todonotes}
\usepackage{comment}
\usepackage{braket}
\usepackage[unicode=true,
 bookmarks=false,
 breaklinks=false,pdfborder={0 0 1},backref=section,colorlinks=false]
 {hyperref}
 \hypersetup{
     colorlinks=true,
     linkcolor=blue,
     filecolor=blue,
     citecolor = magenta,      
     urlcolor=black,
}

\makeatletter

\usepackage{amsthm}\usepackage{amsfonts}\usepackage{times}\usepackage{color}\usepackage{url}

\newcommand{\couic}[1]{}
\newcommand{\couicfootnote}[1]{}
\newcommand{\couicefootnote}[1]{}

\makeatother

\begin{document}

\title{Dirac quantum walk on tetrahedra}

\author{Ugo Nzongani}
\email{ugo.nzongani@universite-paris-saclay.fr}
\affiliation{Universit{\'e} Paris-Saclay, INRIA, CNRS, ENS Paris-Saclay, LMF, 91190 Gif-sur-Yvette, France}

\author{Nathanaël Eon}
\affiliation{Aix-Marseille Université, Université de Toulon, CNRS, LIS, Marseille, France}

\author{Iván Márquez-Martín}
\affiliation{Aix-Marseille Université, Université de Toulon, CNRS, LIS, Marseille, France}

\author{Armando Pérez}
\affiliation{Departamento de Física Teórica and IFIC, Universidad de Valencia and CSIC, Dr. Moliner 50, 46100 Burjassot, Spain}

\author{Giuseppe Di Molfetta}
\email{giuseppe.dimolfetta@lis-lab.fr}
\affiliation{Aix-Marseille Université, Université de Toulon, CNRS, LIS, 13397 Marseille, France}

\author{Pablo Arrighi}
\email{pablo.arrighi@universite-paris-saclay.fr}
\affiliation{Universit{\'e} Paris-Saclay, INRIA, CNRS, ENS Paris-Saclay, LMF, 91190 Gif-sur-Yvette, France}

\begin{abstract}
Discrete-time Quantum Walks (QWs) are transportation models of single
quantum particles over a lattice. Their evolution is driven through
causal and local unitary operators. QWs are a powerful tool for quantum simulation of fundamental physics as some of them have a continuum limit converging to well-known physics partial differential equations, such as the Dirac or the Schrödinger equation. In this work, we show how to recover
the Dirac equation in $(3+1)$--dimensions with a QW evolving in
a tetrahedral space. This paves the way to simulate the Dirac
equation on a curved spacetime. This also suggests an ordered scheme for propagating matter over a spin network, of interest in Loop Quantum Gravity where matter propagation has remained an open problem. 
\end{abstract}


\maketitle

\section{Introduction}
Discrete-time Quantum Walks (QWs) describe situations where a quantum
particle is taking steps on a lattice conditioned on its internal
state, typically a (pseudo) one half spin system. The particle dynamically
explores a large Hilbert space associated with the positions of the
lattice. With QWs, the transport is driven by discrete operations
(shift and coin), which sets it apart from other lattice quantum simulation
frameworks where transport typically rests on hopping between adjacent
sites: all dynamic processes are discrete in space and time. QWs have
been originally introduced by Godoy and Fujita~\cite{godoy92,godoy1993quantum}
as a model of coherent transport over periodic lattices. Similarly
to classical random walk and Markov processes, they have applications
in algorithmics, to express quantum algorithms exhibiting, in some
cases, significant speedup over classical computing~\cite{ambainis2007quantum,roget2020grover,PhysRevResearch.5.033021,slate2021quantum,gamble2010two}.
Most importantly for the present purpose, QWs have an immense track
record for modeling and quantum simulating fundamental physics~\cite{zylberman2022quantum,di2014quantum,de2014quantum,bepari2022quantum,arnault2016quantum,arrighi2014discrete}.
It has been proved that their continuous limit, when it exists, coincides
with one of the two most important equations in physics, namely the Dirac
equation~\cite{arrighi2014dirac,di2012discrete,strauch2006relativistic}
and the Schrödinger equation~\cite{jolly2023twisted}. These results
have made it possible to use the QW as a framework to study the phenomenology
of physical systems, such as Bloch oscillations~\cite{ramasesh2017direct,arnault2020quantum}
or the propagation of matter in strong fields~\cite{di2013quantum,arrighi2016quantum,sajid2019creating}.
Moreover, they have recently proved useful as a fundamental building
block for quantum simulating quantum field theories~\cite{sellapillay2022discrete,eon2022relativistic}
or studying phenomena such as weak ergodicity breaking and Hilbert
space fragmentation~\cite{sellapillay2022entanglement,desaules2022hypergrid}.
QWs have been studied in Bravais lattices in any dimensions, such
as the triangular~\cite{arrighi2018dirac,jay2019dirac} or cubic
grid~\cite{chandrashekar2013two,marquez2017fermion}, and for some
family of regular planar graphs, such as the Apollonian networks~\cite{aristote2020dynamical}.
Most of them have been proved to converge to the Dirac equation in
the space-time continuous limit. For this reason, their use is interesting
in the study and simulation of Dirac materials and topological states~\cite{asboth2012symmetries,verga2018entanglement}.

In this manuscript, we are interested in defining QWs over a tessellation
of Euclidean space by means of tetrahedra. One of the main reasons
why we have taken an interest in this issue is that a tetrahedron
(3-simplex) is the minimal building block of 3D space, similarly
to the role played by a triangular tessellation in a 2D space. The
tessellation that we consider is also the simplest way to fill Euclidean
space with tetrahedra~\cite{sommerville_1922}. As it was shown in
\cite{arrighi2018dirac,jay2019dirac}, the Dirac equation can be simulated
as the continuum limit of a QW defined on a triangular or honeycomb
lattice. This result was further generalized to a curved spacetime
\cite{Arrighi2019}, where the continuum limit also describes the
corresponding Dirac equation, by exploiting the duality between changes
in geometry, and changes in local unitaries that incorporate the deformation
of the metric due to the lack of flatness in spacetime. A crucial
result from this work was that the square lattice fails in reproducing
this duality for arbitrary deformations, in contrast to the triangular
and honeycomb tessellations. From a more practical perspective, the
latter implementations of a QW on a curved geometry need fewer steps
than QWs defined on a rectangular grid \cite{Arnault2017}. All these
qualities make the QW based on tetrahedra a very promising candidate
to investigate the above results for the 3D space.

Finally, another reason for the tetrahedra covering of space is that
it is dual to the 4-valent node, the elementary block of a spin network~\cite{barbieri1998quantum}.
In Loop Quantum Gravity, a spin network is a basis state of the Hilbert
space of the quantized gravitational field on a 3-dimensional hypersurface~\cite{rovelli1995spin}.
Modelling matter propagation over a spin network is still an open
problem~\cite{Bianchi_2013}. In this paper we address this problem
by introducing a toy model where a quantum walker propagates on a
regular spin network. We prove that we recover, in the continuous
space-time limit, the Dirac equation in (3+1)-dimensions.

The organization of the manuscript is the following: in Sec.~\ref{sec:model}
we introduce the formal mathematical model of the QW over tetrahedra.
In Sec. \ref{sec:The-Dirac-equation} we briefly recall the basics
of the Dirac equation and its relation to the Weyl equation. In Sec.~\ref{sec:dirac}
we show how to simulate the Dirac equation
in $3+1$ dimensions on a tetrahedral space. We also discuss how this provides a scheme whereby matter propagates over a spin network. We summarize our main results
in Sec.~\ref{sec:conclusion}. Appendix \ref{app:robust} provides
a robust generalisation to surfaces with boundaries, but only at
the cost of doubling the number of degrees of freedom. We work in
units for which $\hbar=c=1$.

\section{Tetrahedral quantum walk}\label{sec:model}



\begin{figure}
\includegraphics[width=1\columnwidth]{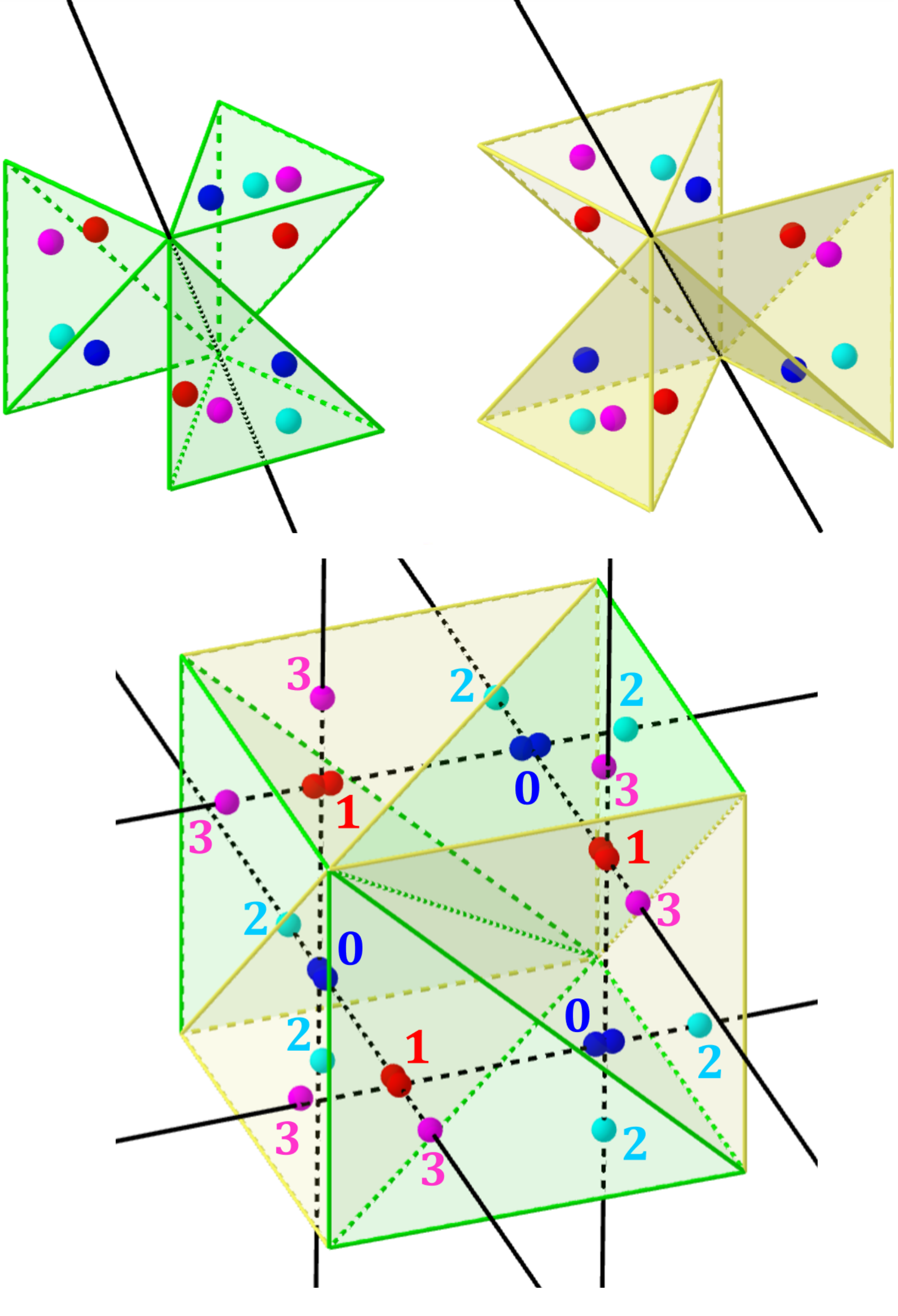}
\caption{(Top) Left-handed and right-handed tetrahedra. (Bottom) Dissection of a cube into six tetrahedra. Coloured dots \textit{blue}, \textit{red}, \textit{cyan} and \textit{magenta} correspond to the amplitudes lying on the facets $i=0,1,2,3$ respectively.}
\label{Fig:filling_cube} 
\end{figure}

\begin{figure}[h!]
\includegraphics[width=0.6\columnwidth]{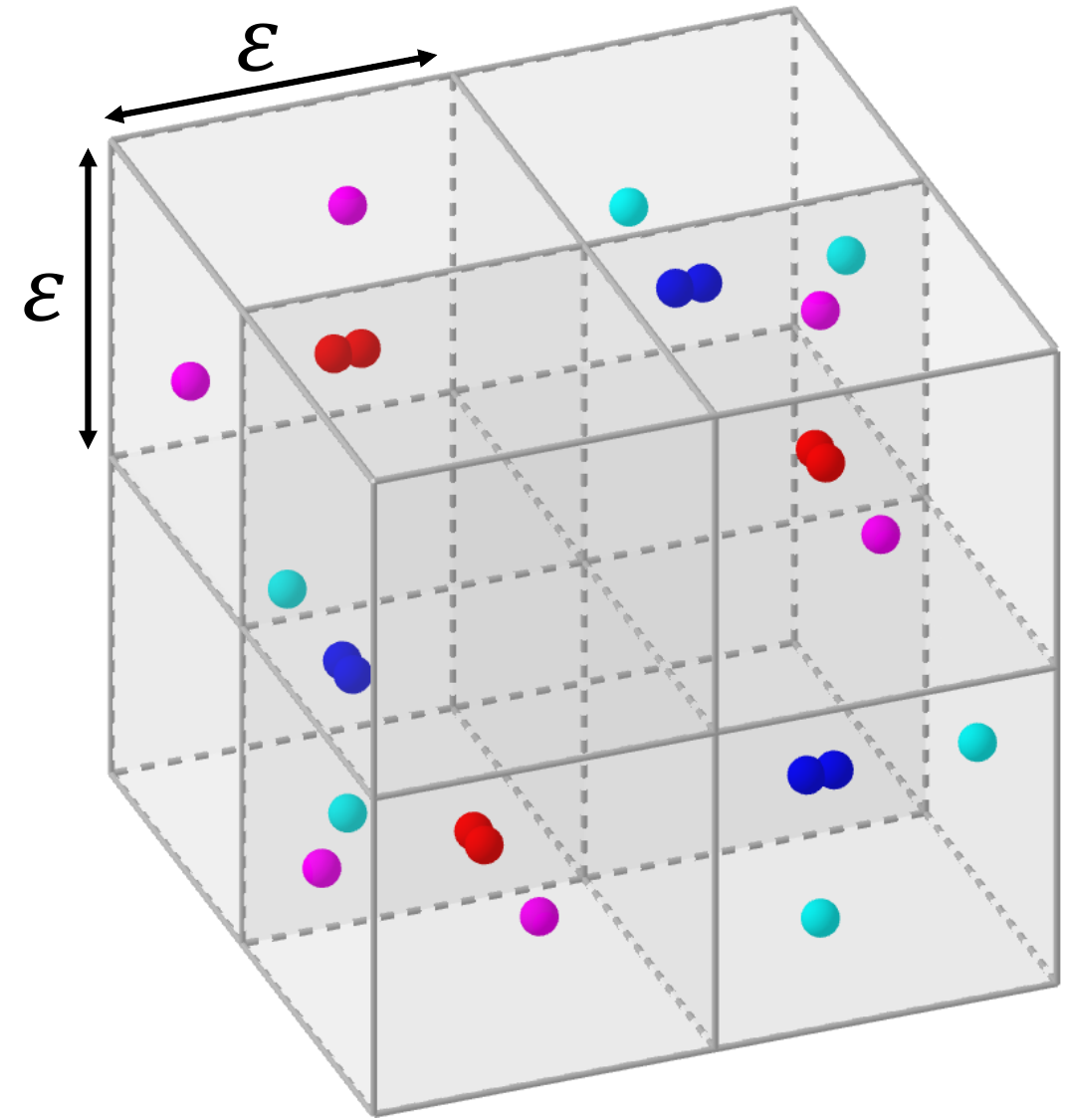}
\caption{Dissection of a cube with 8 smaller cubes.}
\label{Fig:cubic_decomposition} 
\end{figure}

Among all known ways of filling Euclidean space with tetrahedra~\cite{sommerville_1922},
the simplest (albeit non-trivial) way is to dissect a cube into six
3-orthoschemes, i.e. tetrahedra where all four faces are right triangles.
All of them have identical size. However, three of them are left-handed
(LH) and the three others are right-handed (RH) (i.e. mirror to each
other, since we need one LH and one RH to form a cube face) as shown
in Fig.~\ref{Fig:filling_cube}. At time step $t$, each tetrahedron $k$ hosts a $\mathbb{C}^{4}$
vector: 
\begin{equation}
\phi(t,k)=(\phi(t,k,i))_{i=0\dots3}^{T}\quad,\label{eq:deftetra}
\end{equation}
with each complex amplitude $(\phi(t,k,i))_{i=0\dots3}^{T}$ located
on a different facet of the tetrahedron, as represented in Fig.~\ref{Fig:filling_cube}
by coloured dots: \textit{blue}, \textit{red}, \textit{cyan} and \textit{magenta}
corresponding to facets $i=0,1,2,3$ respectively. Assuming there
are no boundaries, each facet is shared by two tetrahedra, a $\mathbb{C}^{2}$
vector on each shared facet can be defined. In the following we focus
on the regular lattice filling up 3D space with such cubes, unrotated\footnote{Thus, cube-cube glueings are RH-RH and LH-LH glueings. Our attempts to systematically enforce RH-LH alternation do not allow us to implement an $x\to y\to z$ sequence of displacements.}, as shown in Fig. \ref{Fig:cube_walk}. This allows
us to adopt the convenient convention that a LH (resp. RH) tetrahedron
be identified in space by means of the coordinates $k=(k_{x},k_{y},k_{z})^{T}$
of its blue $i=0$ (resp. red $i=1$) component with $k\in\varepsilon\mathbb{Z}^{3}$,
with $\varepsilon$ the characteristic lattice spacing which will
be defined below.
Notice that, as shown in Fig. \ref{Fig:cubic_decomposition}, a cube composed of 6 tetrahedra can be further dissected into 8 smaller and identical cubes. The complex amplitudes are then all located in the middle of the small cubes and at the center of their faces. Let $\varepsilon$ be the size of such small cubes. Thus, in a given tetrahedron the distance between the blue and red components is $\varepsilon$, and the one separating a blue (resp. red) from a cyan (resp. magenta) is $\varepsilon/2$.

Dirac QWs over the cubic lattice are of course well-known, and our strategy here will be to fall back on our feet and recover such a known scheme, in spite of the fact that the basic constituents of the lattice are the tetrahedra that make up the cubes and not the cubes themselves. This further subdivision does introduce a number of difficulties in terms of laying out the spinors and orchestrating their propagation, but more fundamentally also in terms of locality: each local unitary gate that makes up the QW needs be local to an individual tetrahedron, or to the joint facet of two glued tetrahedra. The Tetrahedral QW proposed aims to achieve a massless Dirac QW first. Later in the paper we add the mass term to obtain a Dirac QW, again in a way that respects the refined locality constraints of the tetrahedral tesselation.

Across the paper, we say that a quantum operation is
causal when it is allowed to push information from one
tetrahedra to its neighbour, and so on, forming a chain. We say that it
is localized if it decomposes into local operations. We distinguish, however, three types of local operations. Those acting solely on a tetrahedron will be referred to as strictly local. Those acting solely on a facet will be referred to as facet-local. Those allowed to access both the internal amplitudes of a tetrahedron and some amplitudes across its facets will be referred to as weakly local.

\subsection{Shift operator}

\begin{figure}[h!]
\includegraphics[width=0.9\columnwidth]{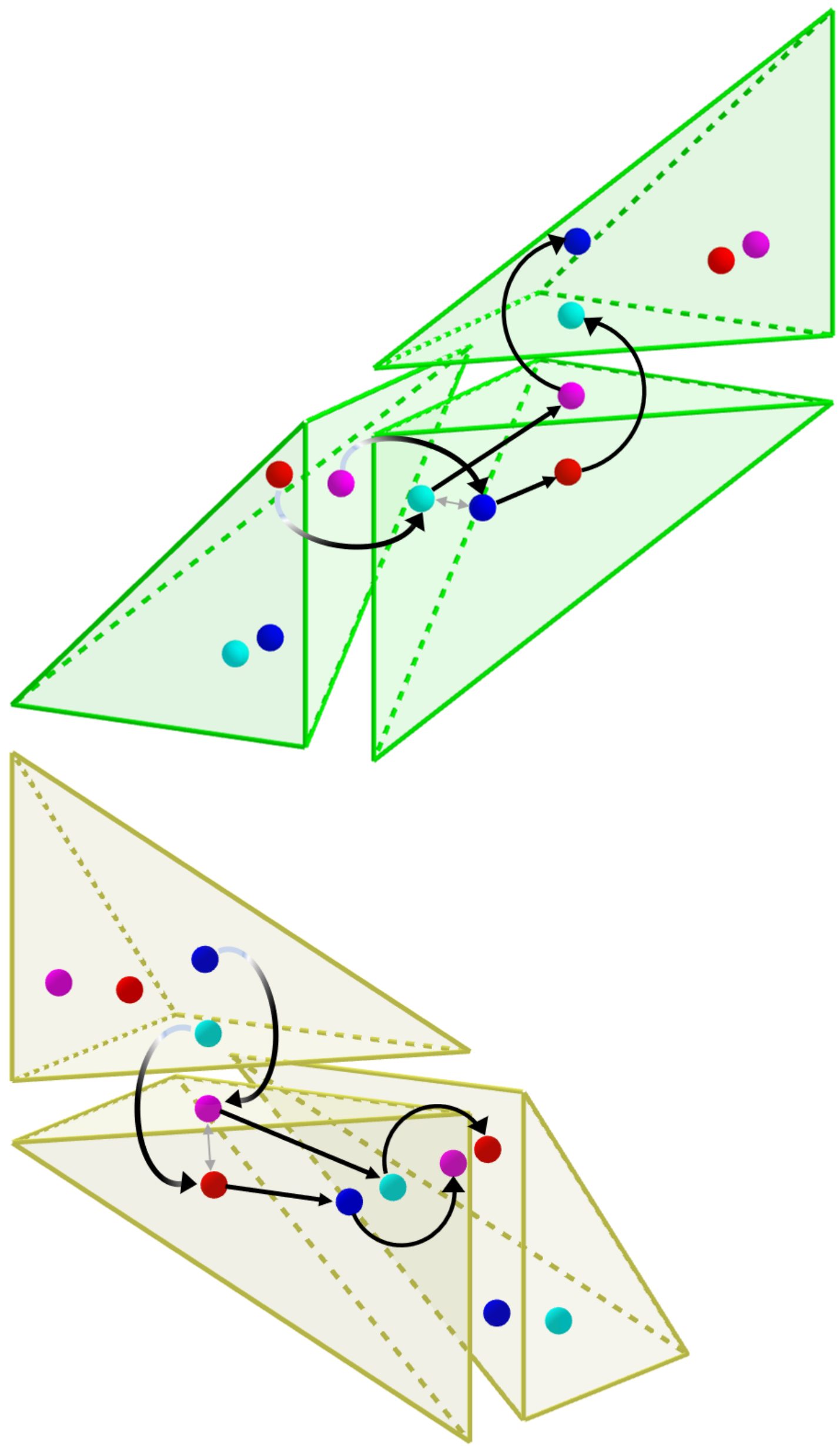}
\caption{Shifting dynamics of the tetrahedral QW working in two steps. First the black arrows act causally on the tetrahedra, with the gradient effect illustrating the need to chain them up. Second the grey arrows (thin ones) that act in a strictly localized manner. Notice that only the coin step shuffles the amplitudes between LH and RH tetrahedra.}
\label{Fig:local_dynamics} 
\end{figure}

The shift operator $S$ decomposes in two steps defined by $S_{B}$
and $S_{G}$. $S_{B}$ is causal: for the LH (resp. RH) (i) the blue and
cyan (resp. red and magenta) components go into the red and magenta
(resp. blue and cyan) facets. At the same time, the red (resp. blue) moves to the cyan (resp. magenta) that lies across the magenta (resp. cyan) facet. Synchronously, the magenta (resp. cyan) moves to the blue (resp. red) of its neighboring tetrahedron. 
$S_{G}$ is
localized, i.e. it is implemented by strictly local operations on each tetrahedron: (ii) the blue and cyan (resp. red and magenta) components are swapped. The first and second steps correspond, respectively, to the black and grey arrows shown in Fig.~\ref{Fig:local_dynamics}.
Let $n(k,i)$ be the neighbor of tetrahedron $k$ on facet $i$. The shifting dynamics therefore reads:
\begin{equation}
    \phi(t+\varepsilon,k)=(S_GS_B\phi)(t,k),
\end{equation}
with
\begin{equation}
    \begin{split}
        (S_B\phi)(t,k) &=
        \begin{cases}
            \begin{pmatrix}
            \phi(t,n(k,2),3) \\
            \phi(t,k,0) \\
            \phi(t,n(k,2),1) \\
            \phi(t,k,2)
            \end{pmatrix} \mbox{ \mbox{ if $k$ is LH}} \\
            \begin{pmatrix}
            \phi(t,k,1) \\
            \phi(t,n(k,3),2) \\
            \phi(t,k,3) \\
            \phi(t,n(k,3),0) 
            \end{pmatrix} \mbox{ otherwise} \\
        \end{cases}, \\\\
        (S_G\phi)(t,k) &=
        \begin{cases}
            \begin{pmatrix}
            \phi(t,k,2) \\
            \phi(t,k,1) \\
            \phi(t,k,0) \\
            \phi(t,k,3)
            \end{pmatrix} \mbox{ if $k$ is LH} \\
             \begin{pmatrix}
            \phi(t,k,0) \\
            \phi(t,k,3) \\
            \phi(t,k,2) \\
            \phi(t,k,1)
            \end{pmatrix} \mbox{ otherwise}
        \end{cases}.
    \end{split}
\end{equation}

\begin{figure*}
     \centering
     \begin{subfigure}[b]{0.8\textwidth}
         \centering
         \includegraphics[width=\textwidth]{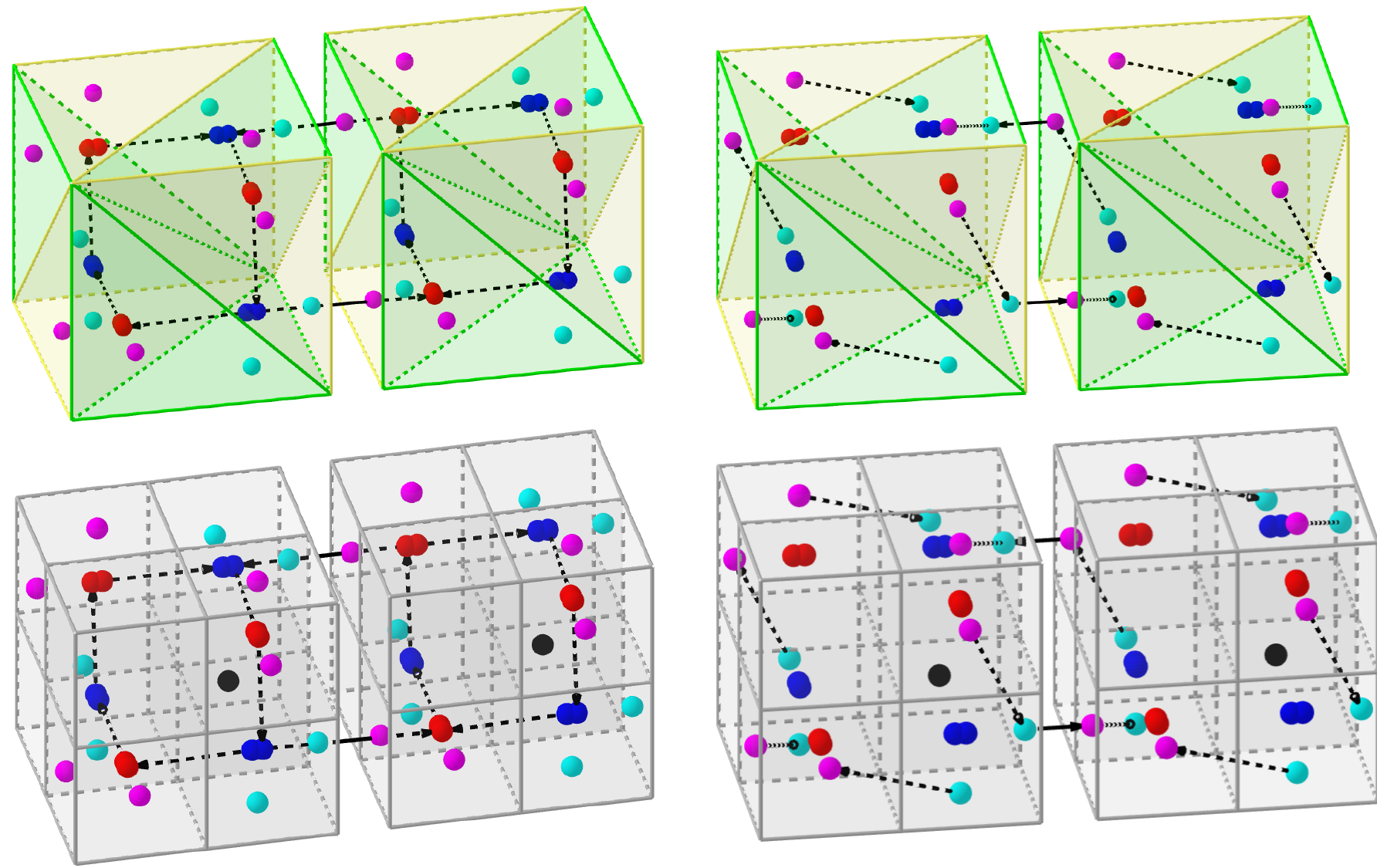}
         \caption{The upper left (resp. right) figure show the global dynamics of the first (resp. second) walker. On the lower figures, we show the same dynamics but replacing the tetrahedra by small cubes. The origin coordinate $k$ corresponds to the black dot located at the bottom right of the small cubes.}
    \label{Fig:cube_walk} 
     \end{subfigure}
     \hfill
     \begin{subfigure}[b]{0.7\textwidth}
         \centering
         \includegraphics[width=\textwidth]{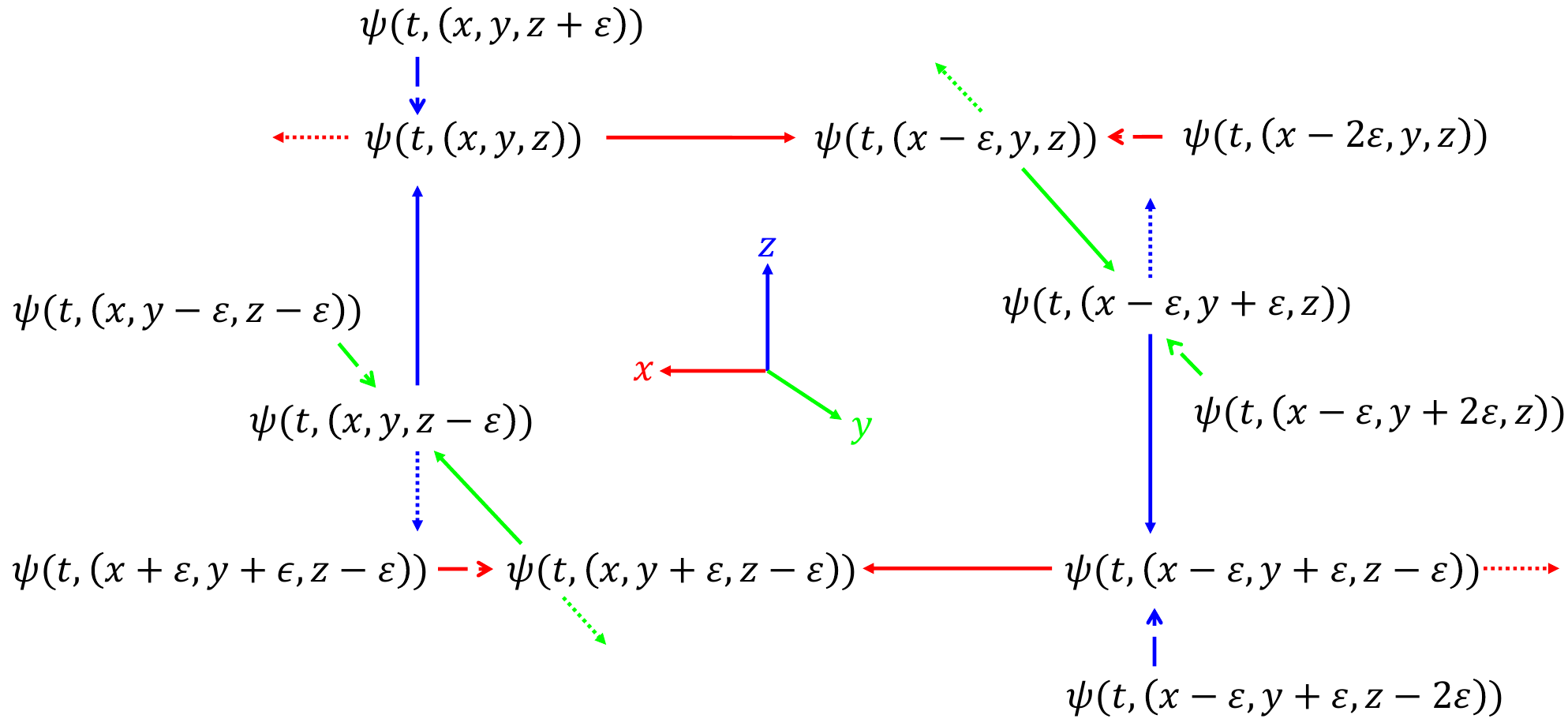}
         \caption{Trajectories of the first walker that moves on a cubic grid.}
         \label{Fig:arrows}
     \end{subfigure}
        \caption{Global dynamics of the tetrahedral QW and trajectories of the first walker.}
        \label{fig:global}
\end{figure*}

As we show on Fig. \ref{Fig:arrows}, the global effect of this dynamics corresponds to the action of a single operator $S$:
\begin{equation}
    \phi(t+\varepsilon,k)=(S\phi)(t,k),
\end{equation}
with
\begin{equation}\label{eq:S}
    (S\phi)(t,k) =
        \begin{cases}
            \begin{pmatrix}
            \phi(t,n(k,2),1) \\
            \phi(t,k,0) \\
            \phi(t,n(k,2),3) \\
            \phi(t,k,2)
            \end{pmatrix} \mbox{ if $k$ is LH} \\
             \begin{pmatrix}
            \phi(t,k,1) \\
            \phi(t,n(k,3),0) \\
            \phi(t,k,3) \\
            \phi(t,n(k,3),2)
            \end{pmatrix} \mbox{ otherwise}
        \end{cases}.
\end{equation}

Notice that the blue and red components are moving along the cubic
grid of Fig.\ \ref{Fig:cubic_decomposition} avoiding the forbidden
positions. By inspecting at Fig.\ \ref{Fig:arrows}, we see that
the cyan and magenta components are following them closely behind,
as if the cyan (resp. magenta) component were linked to the blue (resp.
red). These two walkers are independent of each other. They cross
each other but do not interact. In other words, as proved in Appendix
\ref{app:walk}, we have here two independent walkers evolving according
to the same dynamics, the first one on the blue and red components,
and the second one on the cyan and magenta components.

Let us focus on the first walker. The two-dimensional wave function
living on the shared facet of neighboring tetrahedra are defined as
follows: 
\begin{equation}
    \psi(t,k) = 
    \begin{pmatrix}
        \psi^{\uparrow}(t,k) \\
        \psi^{\downarrow}(t,k)
    \end{pmatrix},
\end{equation}
where the up (resp. down) component lies in the RH (resp. LH) tetrahedra. Let us focus on what happens at coordinate $k$ for instance:
\begin{equation}\label{eq:movement2}
    \psi(t+\varepsilon,k) =
    \begin{pmatrix}
            \psi^{\uparrow}(t,k + \varepsilon u_z) \\
            \psi^{\downarrow}(t,k - \varepsilon u_z)
    \end{pmatrix},
\end{equation}
where $u_z$ is the Bloch vector of $\sigma_z$ i.e. the unit vector along the $z$-axis. It is clear from Eq.\ \eqref{eq:movement2} that the up (resp. down) component is going down (resp. up) in the 3-coordinate which for Fig. \ref{Fig:cube_walk} is vertical. Along one axis, say $z$, we see that the above finite difference equations converge in the continuous limit $\varepsilon\rightarrow{0}$ to the transport equation:
\begin{equation}\label{eq:movement}
    (\partial_t \psi)(t,k) = (\sigma_z\partial_z \psi)(t,k).
\end{equation}
Furthermore, as the axis shown in the center of Fig.\ \ref{Fig:cube_walk} are the one we use and by looking at the direction of the black arrows in Fig. \ref{Fig:arrows}, we see that only black arrows that goes up (resp. down) belongs to an LH (resp. RH) tetrahedron. Thus, the upper (resp. lower) component of $\psi(t,k)$ is located in the RH (resp. LH) tetrahedra.

The direction in which the components moves depends on the coordinates, (again by looking at Fig. \ref{Fig:cube_walk}):
\begin{equation}\label{eq:coord}
    u(k)=
    \begin{cases}
        u_x \mbox{ if } k=\varepsilon(even,odd,odd)^T \mbox{ or } \varepsilon(odd,even,even)^T \\
        u_y \mbox{ if } k=\varepsilon(even,odd,even)^T \mbox{ or } \varepsilon(odd,even,odd)^T \\
        u_z \mbox{ if } k=\varepsilon(even,even,odd)^T \mbox{ or } \varepsilon(odd,odd,even)^T \\
        \varnothing \mbox{ if } k=\varepsilon(even,even,even)^T \mbox{ or } \varepsilon(odd,odd,odd)^T
    \end{cases},
\end{equation}

where $u_\mu$ is the Bloch vector of $\sigma_\mu$ e.g. $u_x=(1,0,0)^T$. Inspecting Fig. \ref{Fig:arrows} we see that the $z$-axis movers become $x$-axis movers. In general, $\mu$-axis movers become $(\mu+1)$-axis movers, as made formal by $u_{\mu+1}=Pu_\mu$ with:
\begin{equation}
    P = \begin{pmatrix}
        0 & 0 & 1 \\ 1 & 0 & 0 \\ 0 & 1 & 0
    \end{pmatrix}.
\end{equation}

The shift evolution of the walker then reads:
\begin{equation}\label{eq:psi}
    \begin{split}
        \psi(t+\varepsilon,k) &=
        \hat{S}
        \begin{pmatrix}
        \psi^{\uparrow}(t,k) \\
        \psi^{\downarrow}(t,k)
    \end{pmatrix} \\
    &=
    \begin{pmatrix}
        \psi^{\uparrow}(t,k+\varepsilon u(k)) \\
        \psi^{\downarrow}(t,k-\varepsilon u(k))
    \end{pmatrix} \\
    &=
    \begin{pmatrix}
        \tau_{u(k),\varepsilon}\psi^{\uparrow}(t,k) \\
        \tau_{u(k),-\varepsilon}\psi^{\downarrow}(t,k)
    \end{pmatrix},
    \end{split}
\end{equation}
  where $\hat{S}$ denotes the effect of $S$ but on the two-dimensional
$\psi$'s, and $\tau_{\mu,\varepsilon}$ is the translation by $\varepsilon$
along the $\mu$-axis. This looks like it does implement the correct
usual partial shift $T_{u,\varepsilon}$ in the $u(k)$ direction
of a cubic QW. Let us look at this in more detail. On a cubic lattice,
a partial shift in some direction $u$ is written as

\begin{equation}
\begin{split}T_{u,\varepsilon}\begin{pmatrix}\psi^{\uparrow}(t,k)\\
\psi^{\downarrow}(t,k)
\end{pmatrix}=\begin{pmatrix}\psi^{\uparrow}(t,k+\varepsilon u)\\
\psi^{\downarrow}(t,k-\varepsilon u)
\end{pmatrix}=\begin{pmatrix}\tau_{u,\varepsilon}\psi^{\uparrow}(t,k)\\
\tau_{u,-\varepsilon}\psi^{\downarrow}(t,k)
\end{pmatrix}.\end{split}
\label{eq:shift}
\end{equation}

\subsection{Coin operator}

Since the walker changes direction along the dynamics, and since our aim is to mimic a transport equation such as $\partial_t\psi =\sigma_x\partial_x\psi+\sigma_y\partial_y\psi+\sigma_z\partial_z\psi$, we have to introduce basis changes to map $\sigma_{\mu-1}$ into $\sigma_\mu$. We notice that $R_{\sigma_z}(\theta)R_{\sigma_x}(\theta)$ with $\theta=\pi/2$ maps the Bloch vector of $\sigma_{\mu-1}$ into that of $\sigma_{\mu}$ with $\mu=1,2,3$, i.e:
\begin{equation}
     \sigma_{\mu} = R_{\sigma_z}(\theta)R_{\sigma_x}(\theta)\sigma_{\mu-1} R^{\dagger}_{\sigma_x}(\theta)R^{\dagger}_{\sigma_z}(\theta)
\end{equation}
since $(R_{\sigma_z}(\theta)R_{\sigma_x}(\theta))^3=-\mathbb{I}$ and we want it to be the identity, we define $C=e^{\mathrm{i}\frac{\pi}{3}}R_{\sigma_z}(\theta)R_{\sigma_x}(\theta)$. This $C$ operation is applied on each facet at once, it is therefore facet-localized.

\subsection{Complete dynamics}

\begin{figure}
\begin{tikzpicture}[scale=0.15]
\tikzstyle{every node}+=[inner sep=0pt]
\draw [red] (21,-10.1) circle (3);
\draw [red] (21,-10.1) circle (2.4);
\draw [blue] (21,-23.6) circle (3);
\draw [red] (21,-36.7) circle (3);
\draw [red] (21,-36.7) circle (2.4);
\draw [blue] (36.1,-36.7) circle (3);
\draw [red] (36.1,-23.6) circle (3);
\draw [red] (36.1,-23.6) circle (2.4);
\draw [blue] (36.1,-10.1) circle (3);
\draw [red] (51.1,-10.1) circle (3);
\draw [red] (51.1,-10.1) circle (2.4);
\draw [blue] (51.1,-23.6) circle (3);
\draw [red] (51.1,-36.7) circle (3);
\draw [red] (51.1,-36.7) circle (2.4);
\draw [gray] (23.528,-8.499) arc (115.08508:64.91492:11.846);
\fill [gray] (33.57,-8.5) -- (33.06,-7.71) -- (32.64,-8.61);
\draw [black] (38.503,-8.321) arc (118.47927:61.52073:10.689);
\fill [black] (48.7,-8.32) -- (48.23,-7.5) -- (47.76,-8.38);
\draw [black] (52.645,-12.66) arc (23.08067:-23.08067:10.689);
\fill [black] (52.65,-21.04) -- (53.42,-20.5) -- (52.5,-20.11);
\draw [gray] (49.66,-20.978) arc (-158.77044:-201.22956:11.4);
\fill [gray] (49.66,-12.72) -- (48.9,-13.29) -- (49.84,-13.65);
\draw [gray] (48.254,-11.039) arc (-76.16681:-103.83319:19.464);
\fill [gray] (38.95,-11.04) -- (39.6,-11.72) -- (39.84,-10.74);
\draw [black] (33.316,-11.207) arc (-73.45184:-106.54816:16.733);
\fill [black] (23.78,-11.21) -- (24.41,-11.91) -- (24.69,-10.96);
\draw [gray] (19.455,-21.04) arc (-156.92857:-203.07143:10.692);
\fill [gray] (19.46,-12.66) -- (18.68,-13.2) -- (19.6,-13.59);
\draw [black] (22.345,-12.773) arc (19.62435:-19.62435:12.139);
\fill [black] (22.34,-20.93) -- (23.08,-20.34) -- (22.14,-20.01);
\draw [black] (19.469,-34.132) arc (-157.46053:-202.53947:10.389);
\fill [black] (19.47,-26.17) -- (18.7,-26.71) -- (19.62,-27.1);
\draw [gray] (22.526,-26.171) arc (22.44958:-22.44958:10.42);
\fill [gray] (22.53,-34.13) -- (23.29,-33.58) -- (22.37,-33.2);
\draw [black] (33.53,-25.133) arc (-66.16844:-113.83156:12.324);
\fill [black] (33.53,-25.13) -- (32.6,-25) -- (33,-25.91);
\draw [gray] (23.672,-22.248) arc (110.63267:69.36733:13.844);
\fill [gray] (23.67,-22.25) -- (24.6,-22.43) -- (24.24,-21.5);
\draw [black] (38.692,-22.104) arc (113.12265:66.87735:12.498);
\fill [black] (38.69,-22.1) -- (39.62,-22.25) -- (39.23,-21.33);
\draw [gray] (48.56,-25.182) arc (-65.31278:-114.68722:11.877);
\fill [gray] (48.56,-25.18) -- (47.62,-25.06) -- (48.04,-25.97);
\draw [black] (49.675,-34.07) arc (-159.29995:-200.70005:11.091);
\fill [black] (49.68,-26.23) -- (48.92,-26.8) -- (49.86,-27.15);
\draw [gray] (52.52,-26.232) arc (20.61792:-20.61792:11.125);
\fill [gray] (52.52,-34.07) -- (53.27,-33.49) -- (52.33,-33.14);
\draw [gray] (38.713,-35.24) arc (112.4806:67.5194:12.782);
\fill [gray] (38.71,-35.24) -- (39.64,-35.4) -- (39.26,-34.47);
\draw [black] (48.434,-38.062) arc (-69.24093:-110.75907:13.638);
\fill [black] (48.43,-38.06) -- (47.51,-37.88) -- (47.86,-38.81);
\draw [gray] (34.735,-34.038) arc (-160.31074:-199.68926:11.539);
\fill [gray] (34.74,-26.26) -- (34,-26.85) -- (34.94,-27.18);
\draw [black] (37.497,-26.245) arc (20.22915:-20.22915:11.293);
\fill [black] (37.5,-34.06) -- (38.24,-33.48) -- (37.3,-33.13);
\draw [black] (23.664,-35.334) arc (110.88171:69.11829:13.708);
\fill [black] (23.66,-35.33) -- (24.59,-35.52) -- (24.23,-34.58);
\draw [gray] (33.449,-38.092) arc (-68.67626:-111.32374:13.473);
\fill [gray] (33.45,-38.09) -- (32.52,-37.92) -- (32.89,-38.85);
\draw [black] (34.823,-20.893) arc (-161.49484:-198.50516:12.738);
\fill [black] (34.82,-12.81) -- (34.1,-13.41) -- (35.04,-13.72);
\draw [gray] (37.56,-12.711) arc (21.58208:-21.58208:11.254);
\fill [gray] (37.56,-20.99) -- (38.32,-20.43) -- (37.39,-20.06);
\end{tikzpicture}
\caption{Independent sub-walks of a 2-dimensional walk. Forget about grey arrows for now. As the amplitudes that were at blue positions follow the black arrows, they now find themselves at red positions, without ever interacting with those which where red initially, who have now left to blue positions.}
\label{fig:2DDecoupling}
\end{figure}
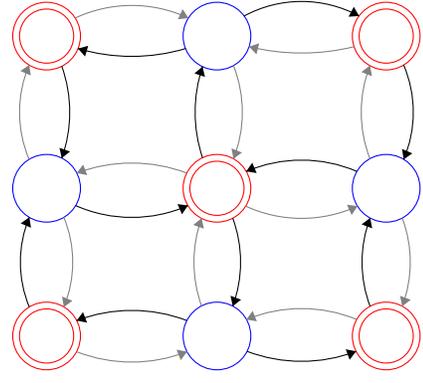

It is important to observe that, within the first quantum walker whose propagation is described by Fig.~\ref{Fig:arrows}, there exist in fact three sub-walkers that are ``independent'', in the sense that they avoid each other and never interact. This is easier to see in 2 spatial dimensions, see Fig.~\ref{fig:2DDecoupling}, where the equivalent of $\hat{S}$ features two independent sub-walkers. If we remove one of the two sub-walkers, we see that the one remaining follows operator $T_x$, then $T_y$, etc. This is why $C\hat{S}$ is a suitable implementation of $CT_yCT_x$.

\section{The Dirac equation\label{sec:The-Dirac-equation}}

Let us briefly recall some properties of the Dirac equation \cite{Peskin:1995ev},
in order to better understand the connection with the QW defined above.
The Dirac equation describes the motion of a relativistic spin $1/2$
particle with mass $m$. It can be written as 
\begin{equation}
{\displaystyle (i\gamma^{\mu}\partial_{\mu}-m)\psi=0,}\label{eq:Diracm}
\end{equation}

where $\gamma^{\mu}$ ($\mu=0,1,2,3)$ are the gamma matrices. This
equation can be decomposed into a pair of coupled Weyl-like (i.e.,
with and additional mass term) equations \cite{Pal2011} acting on
the first two and last two indices of the original four-component
spinor, i.e. by writing 
\begin{equation}
\psi=\left(\begin{array}{c}
\psi_{R}\\
\psi_{L}
\end{array}\right),
\end{equation}

where $\psi_{L}$ and $\psi_{R}$ are each two-component Weyl spinors.
This can be achieved by using the so-called chiral representation
of the gamma matrices \cite{Peskin:1995ev}. Following this procedure,
Eq. (\ref{eq:Diracm}) becomes equivalent to a pair of coupled Weyl
equations 
\begin{equation}
{\displaystyle i\sigma^{\mu}\partial_{\mu}\psi_{R}=m\psi_{L}}\label{eq:WeylR}
\end{equation}
\begin{equation}
{\displaystyle i{\overline{\sigma}}^{\mu}\partial_{\mu}\psi_{L}=m\psi_{R}}.\label{eq:WeylL}
\end{equation}
In the above equations, $\sigma^{\mu}$ has components ${\displaystyle (I_{2},\sigma^{i})}$
and ${\displaystyle {\bar{\sigma}}^{\mu}}$ has components ${\displaystyle (I_{2},-\sigma^{i})}$,
with $I_{2}$ the two-dimensional identity matrix. In the massless
$m=0$ case both equations become two true decoupled Weyl equations.
The coupling between both equations (\ref{eq:WeylR}) and (\ref{eq:WeylL})
is associated to the action of the $\gamma{^{0}}$ matrix, which ``flips''
both Weyl spinors.

In the following sections we will make use of the above ideas to construct
a QW that describes the massive Dirac equation. First we construct a QW for the massless Dirac equation. Then, the necessary flipping operation will be
introduced to include the mass term.

\section{Dirac Quantum Walk}\label{sec:dirac}

\subsection{Massless Dirac QW}

After multiplication by $\gamma^{0}$, the massless Dirac equation
takes the form 
\begin{equation}
\begin{split}\mathrm{i}\partial_{0}\Psi & =\hat{D}\Psi\quad\textrm{with}\\
\hat{D} & =\mathrm{i}\sum_{j=1}^{3}\alpha^{j}\partial_{j}
\end{split}
\quad,
\end{equation}
where $\alpha^{j}=\gamma^{0}\gamma^{j}$. To obtain the massless Dirac equation, we therefore need a four-dimensional spinor and a new coin acting on these four amplitudes. Let $\Psi(t,k)=(\Psi_i(t,k))^T_{i=0,\dots,3}$ be such a spinor. Its components are the coloured dots located on the small cubes of Fig.\ \ref{Fig:arrows}. Therefore, such a spinor contains components of two tetrahedra, LH and RH. Note that this spinor contains either blue and cyan, or red and magenta components, depending on the parity of the position. For those containing blue and cyan (resp. red and magenta) components; the first component ($i=0$) is the cyan (resp. magenta) dot of the RH tetrahedron, the second ($i=1$) is the blue (resp. red) of the RH, the third ($i=2$) is the blue (resp. red) of the LH, and the last one ($i=3$) is the cyan (resp. magenta) of the LH tetrahedron. It turns out that the chiral representation keeps the strict locality, thus:
\begin{equation}\label{eq:choice_rpz}
    \alpha^j = \sigma_3 \otimes \sigma_j.
\end{equation}
The coin operator $\hat{C}$ must verify $\alpha^\mu=\hat{C}\alpha^{\mu-1} \hat{C}^\dagger$. That way, $\hat{C}=\sigma_0\otimes C$ is the coin operator. We keep identical the first (black) and second (grey) steps of the shift operation, only reexpressing their effects in terms of these spinors:
\begin{equation}\label{eq:partial}
    \begin{split}
    (\mathcal{T}_{B,u(k),\varepsilon}\Psi)(t,k) &=
    \begin{cases}
        \begin{pmatrix}
            \Psi_0(t,k+\varepsilon u(k)) \\
            \Psi_1(t,k+\varepsilon u(k)) \\
            \Psi_3(t,k-\varepsilon u(k)) \\
            \Psi_2(t,k-\varepsilon u(k))
        \end{pmatrix} \mbox{ if $k$ is LH} \\
        \begin{pmatrix}
            \Psi_1(t,k+\varepsilon u(k)) \\
            \Psi_0(t,k+\varepsilon u(k)) \\
            \Psi_2(t,k-\varepsilon u(k)) \\
            \Psi_3(t,k-\varepsilon u(k))
        \end{pmatrix} \mbox{ otherwise}
    \end{cases}, \\
    (\mathcal{T}_{G,u(k),\varepsilon}\Psi)(t,k) &=
    \begin{cases}
        \begin{pmatrix}
            \Psi_0(t,k) \\
            \Psi_1(t,k) \\
            \Psi_3(t,k) \\
            \Psi_2(t,k)
        \end{pmatrix} \mbox{ if $k$ is LH} \\
        \begin{pmatrix}
            \Psi_1(t,k) \\
            \Psi_0(t,k) \\
            \Psi_2(t,k) \\
            \Psi_3(t,k)
        \end{pmatrix} \mbox{ otherwise}
    \end{cases}.
    \end{split}
\end{equation}
Hence,
\begin{equation}
    (\mathcal{T}_{G,b(k),\varepsilon}\mathcal{T}_{B,u(k),\varepsilon}\Psi)(t,k) = \begin{pmatrix}
        \Psi_0(t,k+\varepsilon u(k)) \\
        \Psi_1(t,k+\varepsilon u(k)) \\
        \Psi_2(t,k-\varepsilon u(k)) \\
        \Psi_3(t,k-\varepsilon u(k))
    \end{pmatrix},
\end{equation}
performs a partial shift in the $u(k)$ direction. For the $z$ axis, this approximates as:
\begin{equation}
    \left( \mathbb{I}+\varepsilon (\sigma_3\otimes\sigma_3)\partial_3 \right) \Psi = \mathcal{T}_{G,3,\varepsilon}\mathcal{T}_{B,3,\varepsilon}\Psi+O(\varepsilon^2).
\end{equation}
In the same way, along the $y$-axis,
\begin{equation}
    \left( \mathbb{I}+\varepsilon (\sigma_3\otimes\sigma_2)\partial_2 \right) \Psi = \hat{C}^{\dagger}\mathcal{T}_{G,2,\varepsilon}\mathcal{T}_{B,2,\varepsilon}\hat{C}\Psi+O(\varepsilon^2).
\end{equation}
Lastly, for the $x$-axis we obtain:
\begin{equation}
        \left( \mathbb{I}+\varepsilon (\sigma_3\otimes\sigma_1)\partial_1 \right) \Psi 
        = \hat{C}\mathcal{T}_{G,1,\varepsilon}\mathcal{T}_{B,1,\varepsilon}\hat{C}^{\dagger}\Psi+O(\varepsilon^2).
\end{equation}
The time evolution of the massless Dirac QW thus reads:
\begin{equation}
\Psi(t+\varepsilon,k) = \hat{W}_\varepsilon\Psi(t,k),
\end{equation}
with:
\begin{equation}
    \label{eq:second_weyl}
     \hat{W}_\varepsilon = \hat{C}^\dagger\mathcal{T}_{G,2,\varepsilon}\mathcal{T}_{B,2,\varepsilon}\hat{C}^\dagger\mathcal{T}_{G,1,\varepsilon}\mathcal{T}_{B,1,\varepsilon}\hat{C}^\dagger\mathcal{T}_{G,3,\varepsilon}\mathcal{T}_{B,3,\varepsilon}.
\end{equation}
Using Eq. \eqref{eq:psi2approx} and Eq. \eqref{eq:second_weyl} we get:
\begin{equation}
    \begin{split} 
        \Psi(t+\varepsilon,k) &= \prod_{j=1}^3 \hat{C}^\dagger (\mathbb{I}+\varepsilon(\sigma_3\otimes \sigma_3)\partial_j)\Psi(t,k) + O(\varepsilon^2) \\
        \Psi(t+\varepsilon,k) &= \prod_{j=1}^3 (\mathbb{I}+\varepsilon(\sigma_3\otimes \sigma_j)\partial_j)
        \Psi(t,k) + O(\varepsilon^2) \\
        (\mathbb{I}+\varepsilon\partial_0)\Psi(t,k) &= \left(\mathbb{I}+\sum_{j=1}^3\varepsilon(\sigma_3\otimes \sigma_j)\partial_j\right)\Psi(t,k) + O(\varepsilon^2) \\
        \partial_0\Psi(t,k) &= \sum_{j=1}^3 (\sigma_3\otimes \sigma_j)\partial_j\Psi(t,k)+ O(\varepsilon^2).
    \end{split}
\end{equation}
Therefore, in the continuum limit we approximate the massless Dirac equation in (3+1)-dimensions.

\subsection{Dirac with mass}\label{subsec:diracmass}

\begin{figure}
    \centering
    \includegraphics[width=0.8\columnwidth]{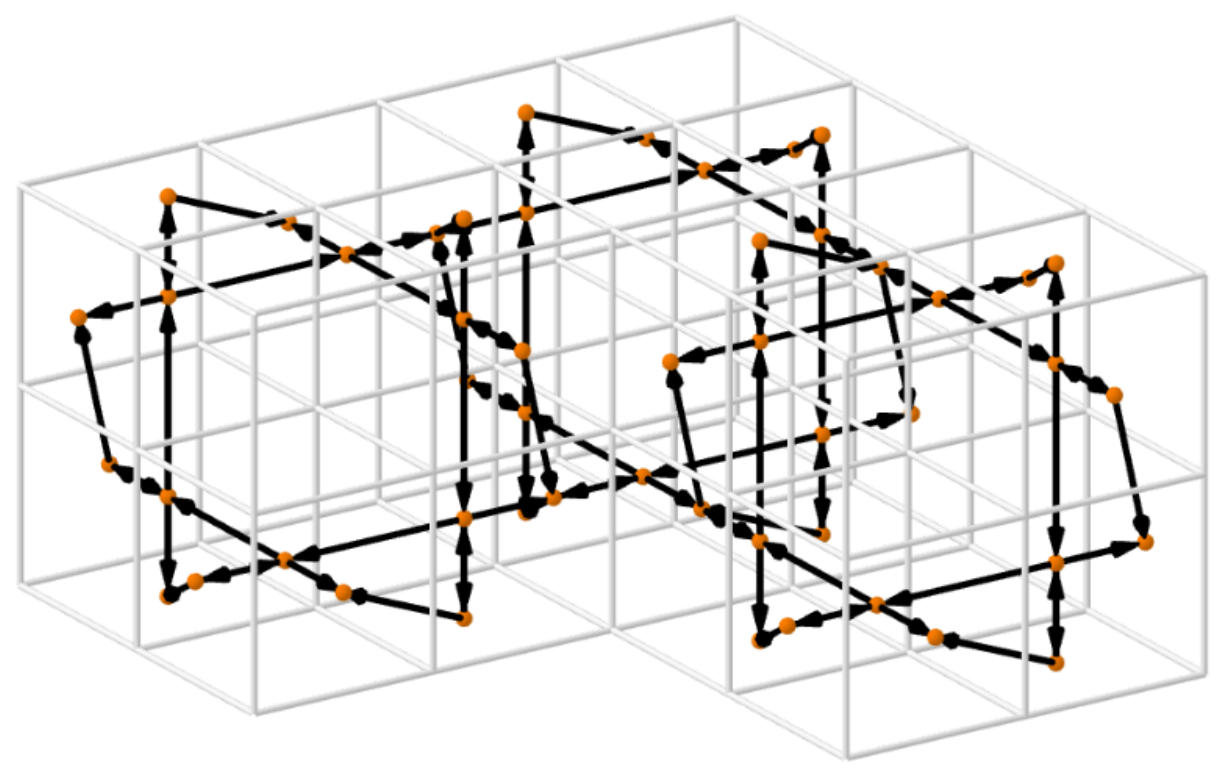}
    \caption{Displacement graph of the tetrahedral Dirac QW.}
    \label{fig:graph}
\end{figure}

The QW we introduced only portrays the movement of massless particles. To prevent them from travelling at the speed of light we introduce
an additional new coin acting as the mass term. The massive Dirac
equation now takes the form: 
\begin{equation}
\begin{split}\mathrm{i}\partial_{0}\Psi & =\mathcal{D}\Psi\quad\textrm{with}\\
\mathcal{D} & =m\gamma^{0}+\mathrm{i}\sum_{j=1}^{3}\alpha^{j}\partial_{j}
\end{split}
\quad,
\end{equation}
where $m$ is the mass of the particle. The $\alpha^j$ matrices are that of Eq. \eqref{eq:choice_rpz} and $\gamma^0$ is defined as: 
\begin{equation}
\gamma^{0}=\sigma_{1}\otimes\sigma_{0}.\label{eq:alpha0}
\end{equation}
Let $M=e^{-\mathrm{i}\varepsilon m\gamma^{0}}$ be the mass coin. 
According to our choice of representation in Eq. \eqref{eq:choice_rpz} and Eq. \eqref{eq:alpha0}, the coin $\hat{C}$ makes the first and second (resp. third and fourth) components of $\Psi$ interact with each others, meaning that it is a strictly localized operator. On the other hand, the mass coin $M$ makes the first and third (resp. second and fourth) components interact with each others, which means that it is weakly localized i.e. it can be decomposed into weakly local operators, each of which is allowed to access an amplitude from across one of the facets of the tetrahedron. \\
Moreover, we notice that:
\begin{equation}\label{eq:no_impact_coins}.
    \begin{split}
        \hat{C}M\hat{C}^\dagger &= M \\
        M\hat{C}M^\dagger & =\hat{C}
    \end{split}\quad ,
\end{equation}
making it is possible to successively apply the coins without disturbing the walk. The time evolution of the massive Dirac tetrahedral QW is defined as:
\begin{equation}
\Psi(t+\varepsilon,k) = (\mathcal{W}_\varepsilon\Psi)(t,k),
\end{equation}
where
\begin{equation}
    \label{eq:dirac_final}
    \begin{split}
         \mathcal{W}_\varepsilon &= \hat{C}^\dagger\mathcal{T}_{G,2,\varepsilon}M\mathcal{T}_{B,2,\varepsilon}\hat{C}^\dagger\mathcal{T}_{G,1,\varepsilon}M\mathcal{T}_{B,1,\varepsilon}\hat{C}^\dagger\mathcal{T}_{G,3,\varepsilon}M\mathcal{T}_{B,3,\varepsilon} \\
         &= \mathcal{\hat{T}}_{G,2,\varepsilon}\mathcal{T}_{B,2,\varepsilon}\mathcal{\hat{T}}_{G,1,\varepsilon}\mathcal{T}_{B,1,\varepsilon}\mathcal{\hat{T}}_{G,3,\varepsilon}\mathcal{T}_{B,3,\varepsilon}
     \end{split},
\end{equation}
with $\mathcal{\hat{T}}_{G,\mu,\varepsilon} = \hat{C}^\dagger\mathcal{T}_{G,\mu,\varepsilon}M$. 

In the end, the tetrahedral Dirac QW is just the repetition of two steps, one being causal and the other one localized. More precisely, according to Eq.\ \eqref{eq:dirac_final} it consists in: (i) a causal shift $\mathcal{T}_{B,\mu,\varepsilon}$ and (ii) applying the mass coin $M$, performing a shift $\mathcal{T}_{W,\mu,\varepsilon}$ and applying a basis change coin $\hat{C}^\dagger$, all three localized.

In the continuum limit we do obtain the massive Dirac equation since:
\begin{equation}
    \begin{split}
        \Psi(t+\varepsilon,k) &= (\mathbb{I}-\mathrm{i}\varepsilon D) \Psi(t,k) + O(\varepsilon^2) \\
        &= \prod_{j=1}^3 \hat{C}^\dagger (\mathbb{I}+\varepsilon(\sigma_3\otimes \sigma_3)\partial_j)M\Psi(t,k) + O(\varepsilon^2) \\
        &= \prod_{j=1}^3 (\mathbb{I}+\varepsilon(\sigma_3\otimes \sigma_j)\partial_j)M\Psi(t,k) +O(\varepsilon^2).
    \end{split}
\end{equation}
Moreover, as $M=\mathbb{I}-\mathrm{i}\varepsilon m\gamma^{0}+O(\varepsilon^{2})$,
we have: 
\begin{equation}
\begin{split}(\mathbb{I}+\varepsilon\partial_{0})\Psi(t,k) & =\left(\mathbb{I}-3\mathrm{i}m\varepsilon\gamma^{0}+\sum_{j=1}^{3}\varepsilon(\sigma_{3}\otimes\sigma_{j})\partial_{j}\right)\Psi(t,k)\\
 & +O(\varepsilon^{2})\\
\partial_{0}\Psi(t,k) & =\left(-3\mathrm{i}m\gamma^{0}+\sum_{j=1}^{3}(\sigma_{3}\otimes\sigma_{j})\partial_{j})\right)\Psi(t,k)\\
 & +O(\varepsilon^{2}).
\end{split}
\label{eq:Diracwith3m}
\end{equation}

Let us notice that it would be possible to modify the QW operator
defined in Eq. (\ref{eq:dirac_final}) so that the mass term appears
only once, by redefining 
\begin{equation}
\mathcal{W}_{\varepsilon}=M\hat{C}^{\dagger}\mathcal{T}_{G,2,\varepsilon}\mathcal{T}_{B,2,\varepsilon}\hat{C}^{\dagger}\mathcal{T}_{G,1,\varepsilon}\mathcal{T}_{B,1,\varepsilon}\hat{C}^{\dagger}\mathcal{T}_{G,3,\varepsilon}\mathcal{T}_{B,3,\varepsilon}.
\end{equation}
It is easy to check that the corresponding continuum limit is the
same as in Eq. (\ref{eq:Diracwith3m}), without the factor $3$ in
front of the first term on the rhs. The latter definition might be
more economical when it comes to implementing the QW on experimental
devices. However, we preferred to use the first definition of $\mathcal{W}_{\varepsilon}$
because it can be more elegantly integrated on the action of each
of the $\mathcal{\hat{T}}_{B,\mu,\varepsilon}$ operators, instead
of an isolated operation that only appears at the end of each time
step.

\subsection{Spin network}\label{subsec:spin}

\begin{figure}[h!]
    \centering
    \includegraphics[width=0.6\columnwidth]{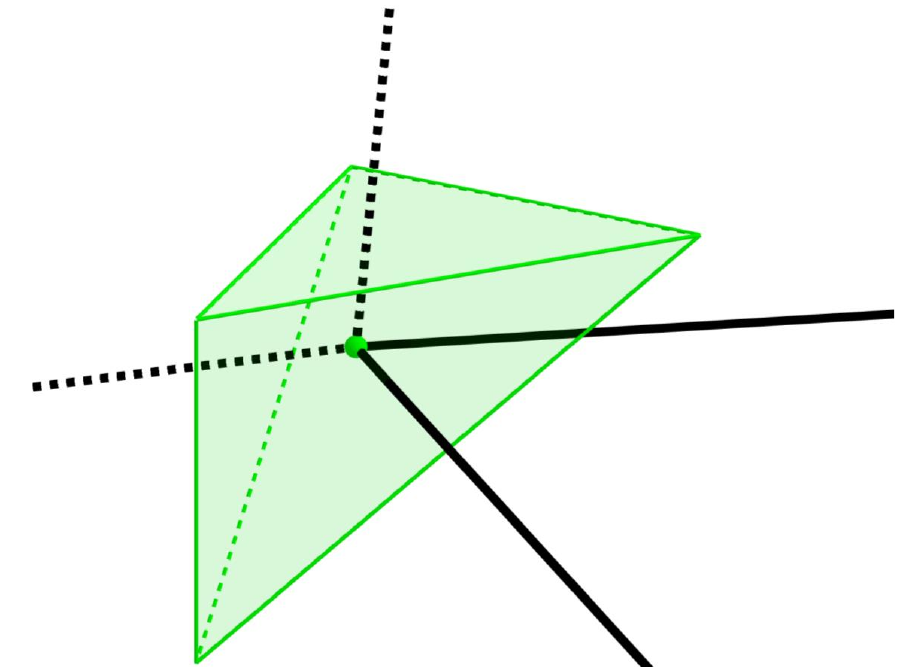}
    \caption{Simplicial complex and dual representation of a LH tetrahedron.}
    \label{fig:2dual}
\end{figure}

\begin{figure*}
     \centering
     \begin{subfigure}[b]{0.4\textwidth}
         \centering
         \includegraphics[width=\textwidth]{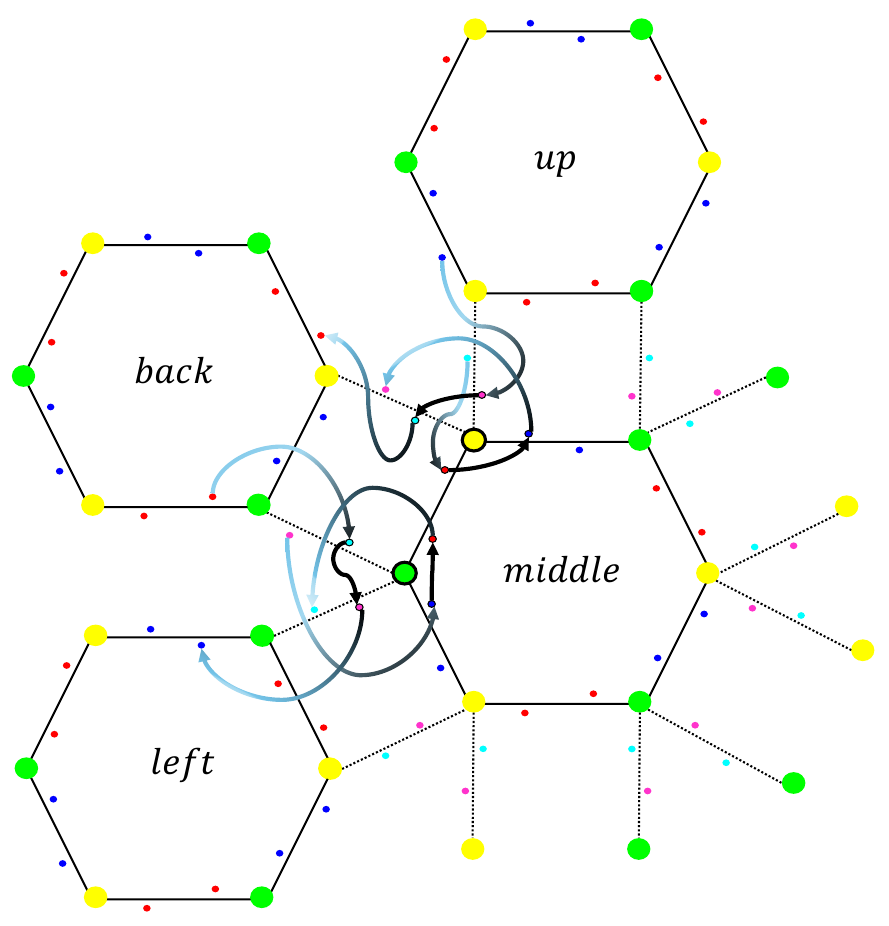}
         \caption{Causality of the black shift, represented with the arrows, which is the first step of the shifting dynamics. Arrows coming out of or going into the observed tetrahedra have a gradient effect to show the need to chained them up.}
    \label{fig:par} 
     \end{subfigure}
     \hfill
     \begin{subfigure}[b]{0.4\textwidth}
         \centering
         \includegraphics[width=\textwidth]{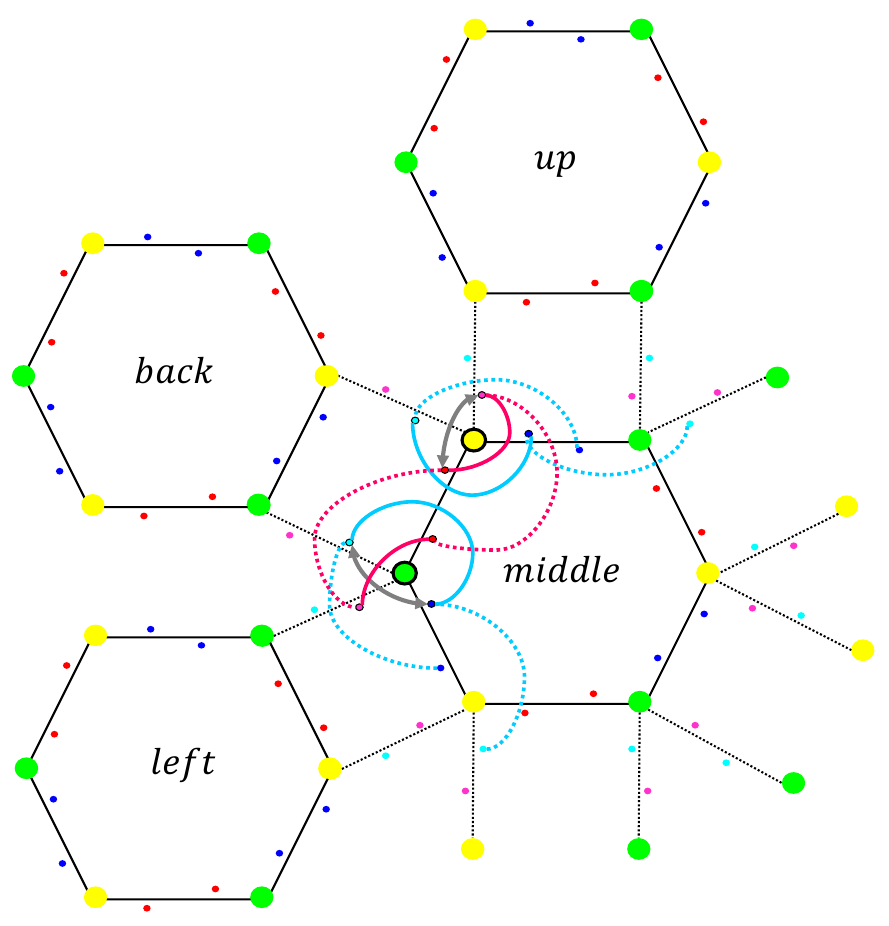}
         \caption{Strict locality of the grey shift and basis coin operator, the shift corresponds to the grey arrows (second step of the shifting dynamics) and the action of the strictly localized coin $\hat{C}$ is represented with the blue and red lines. Weak locality of $M$ is shown by the dashed blue and red lines.}
         \label{fig:pur}
     \end{subfigure}
    \caption{Part of the spin network associated to the Dirac tetrahedral QW. Green and yellow nodes corresponds to the LH and RH tetrahedra, and two nodes are linked if their corresponding tetrahedra share a facet. The amplitudes lie on the edges. Therefore, a cube is represented by an hexagon. The position of the cubes in relation to each other is indicated by the label in their center. We show the action of the shift and coin operators for the two tetrahedra framed in black. In Fig.\ \ref{fig:par} we focus on the causal operations (the black shift), and in Fig.\ \ref{fig:pur} we show the localized ones (grey shift, basis coin and mass coin).}
    \label{fig:dual}
\end{figure*}

In this work we have considered a particular simplicial complex, namely a regular lattice of tetrahedra. It is quite common to consider the dual graph of a simplicial complex, namely by allocating, to each simplex a node, and to each glueing between facets, a link between the nodes, see Fig.~\ref{fig:2dual}. When the simplices are tetrahedra as in our case, the dual graph is a four-valent graph quite close to a spin network, see Fig.\ \ref{fig:dual}.

Spin networks form a basis of the Hilbert space of 3D geometries in Loop Quantum Gravity. The questions of how to propagate matter, such as massive Dirac particles, is a long-standing problem in Loop Quantum Gravity. Our four-valent graphs are variants, the main difference is
that the neighbours of each node are ordered, and that there are left-handed and right-handed nodes, but those extra ingredients seem crucial in order to achieve matter propagation. Indeed, the present paper has just achieved a scheme which, upon the spin network-like graph that is closest to flat space, reproduces the Dirac equation in flat space. We wonder whether there is a sense in which this exact same scheme, generalized to a spin network (or a superposition of them) representing curved space, reproduces the Dirac equation in this curved space.

\section{Summary and perspectives}\label{sec:conclusion}

In this paper we tessellate the 3D space with just two kinds of tetrahedra,
whose shapes are mirror of each other. We place the amplitudes on
their facets, and we alternate two steps of dynamics according to
Eq.\ \eqref{eq:dirac_final}: (i) we do a causal shift,
displacing amplitudes within the tetrahedron and across their facets (ii) we apply a weakly localized mass coin, we do a partial shift
and apply a basis change coin, both strictly localized to each tetrahedron.
We prove that, in the continuum limit, we obtain the Dirac equation
in (3+1)-dimensions. By taking the graph that is dual to the simplicial
complex, we obtain a variant of a Loop Quantum Gravity spin network.
The main difference is that the neighbors of each node are ordered,
and that there are left-handed and right-handed nodes. Our scheme
then provides a way to propagate spin $1/2$ particles over these
modified spin networks. Interestingly, these modifications seem necessary
in order to achieve such matter propagation.

There are many possible avenues to explore. $3+1$ Dirac QW on the
grid have been extended to include the electromagnetic field, or to
account for the multi-particle sector (quantum cellular automata).
One could try to do the same for tetrahedra. More interestingly perhaps,
there exists $3+1$ `curved' Dirac QW on the grid which, in the continuum
limit, yield the Dirac equation in curved spacetime, as modelled by
a metric field, which is used to parametrize the coin. Alternatively,
curved space can be modelled in a combinatorial manner just by glueing
tetrahedra in a way that matches the discretization of 3D manifold.
We wonder whether running our Tetrahedral QW over some discretized
manifold, yields the same behaviour as the $3+1$ curved Dirac QW
over the manifold, in some limit.

\section{Acknowledgements}

This project/publication was made possible through the support of
the ID\# 62312 grant from the John Templeton Foundation, as part of
the \href{https://www.templeton.org/grant/the-quantum-information-structure-of-spacetime-qiss-second-phase}{‘The Quantum Information Structure of Spacetime’ Project (QISS) }.
The opinions expressed in this project/publication are those of the
author(s) and do not necessarily reflect the views of the John Templeton
Foundation. This work is supported by the PEPR integrated project
EPiQ ANR-22-PETQ-0007 part of Plan France 2030, by the ANR JCJC DisQC
ANR-22-CE47-0002-01 founded from the French National Research Agency
and with the support of the french government under the France 2030
investment plan, as part of the Initiative d'Excellence d'Aix-Marseille
Université - A{*}MIDEX AMX-21-RID-011. From A.P. side, this work has
been founded by the Spanish MCIN/AEI/10.13039/501100011033 grant PID2020-113334GB-I00,
Generalitat Valenciana grant CIPROM/2022/66, the Ministry of Economic
Affairs and Digital Transformation of the Spanish Government through
the QUANTUM ENIA project call - QUANTUM SPAIN project, and by the
European Union through the Recovery, Transformation and Resilience
Plan - NextGenerationEU within the framework of the Digital Spain
2026 Agenda, and by the CSIC Interdisciplinary Thematic Platform (PTI+)
on Quantum Technologies (PTI-QTEP+). This project has also received
funding from the European Union’s Horizon 2020 research and innovation
program under grant agreement CaLIGOLA MSCA-2021-SE-01-101086123.

\appendix

\section{Same path}\label{app:walk}

\begin{figure}
    \centering
    \includegraphics[scale=0.35]{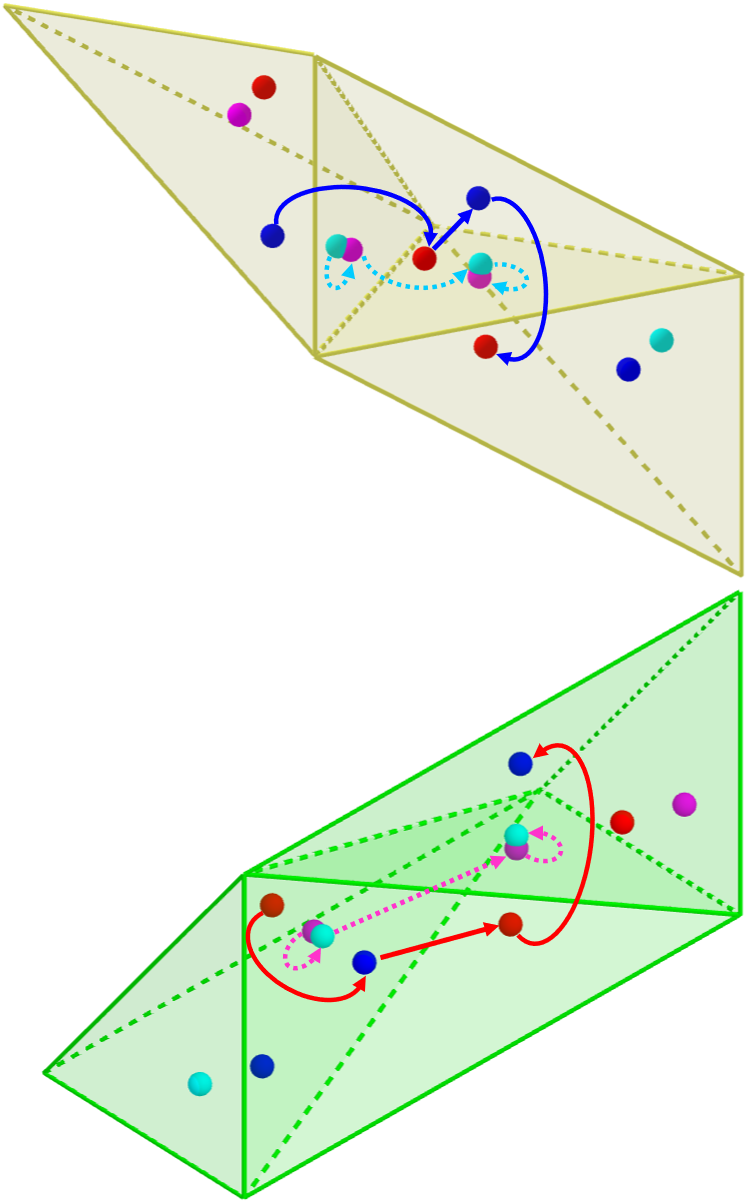}
    \caption{Path followed by amplitudes of the first and second walkers for the LH and RH tetrahedra after 3 time steps. The blue and red (resp. cyan and magenta) arrows are displayed with filled (resp. dashed) linestyle.}
    \label{fig:following_path}
\end{figure}

To show that the second walker follows the steps of the first one, we may compare the coordinates after one time step of a blue or red component to that of a cyan or magenta components initially located in the same tetrahedron. By looking at the direction of the arrows of Fig. \ref{Fig:arrows} we see that in the LH (resp. RH) tetrahedra the direction of motion is positive (resp. negative) with respect to the three axis $x,y,z$ shown in Fig.\ \ref{Fig:cube_walk}. Let $k_{1}(t)$ (resp. $k_{2}(t)$) be the coordinate of a blue or red (resp. cyan or magenta) component of the tetrahedron of coordinate $k$ at time $t$:

\begin{equation}
     k_{1}(t) = \begin{pmatrix}
        k_x \\ k_y \\ k_z
    \end{pmatrix}, k_{2}(t) = k_1(t) \pm \varepsilon
    \begin{pmatrix}
        \kappa_x \\ \kappa_y \\ \kappa_z
    \end{pmatrix}
\end{equation}
where one of the $\kappa_\mu$ is equal to $1/2$ and the rest are $0$. The non-vanishing element is located on the axis on which the first walk receives information at time $t$, as shown in Fig.\ \ref{Fig:cube_walk}. After one time step, since a movement along one of the axis by a distance $\varepsilon$ has been made, it is straightforward noticing that:
\begin{equation}
     k_1(t+\varepsilon) = k_1(t) \pm
    \varepsilon
     \begin{pmatrix}
        \kappa_x \\ \kappa_y \\ \kappa_z
    \end{pmatrix},
\end{equation}
where one of the $\kappa_\mu$ is equal to $1$ and the values of the others are identically $0$. Similarly, by looking at Fig.~\ref{Fig:arrows} we see that:
\begin{equation}
     k_2(t+\varepsilon) = k_2(t) \pm
    \varepsilon
     \begin{pmatrix}
        \kappa_x \\ \kappa_y \\ \kappa_z
    \end{pmatrix}.
\end{equation}
Now, two scenarios may arise. Either two of the $\kappa_\mu$ are 1/2 and the remainder is 1, or the component is simply swaped on the other side of the facet on which it was, and so without involving a change of coordinates. There is only a difference of $\pm \varepsilon/2$ in one of the coordinates of $k_1$ and $k_2$ at time $t$. This proves that the second walker follows the first very closely. We show on Fig. \ref{fig:following_path} that the amplitudes of first and second walkers are following each others. Thus, we consider that they follow the same cubic path.

\section{Robust QW}\label{app:robust}

As explained in the core of the paper, one of the main motivations of this study of QW over tetrahedra is to then be able to apply the same QW scheme over arbitrary graphs of degree bounded by four: whether to represent curvature in the combinatorial languague of simplicial complexes; exotic crystalling structures; boundaries or defects. For instance in \cite{roget2020grover, PhysRevResearch.5.033021} defects are coded as missing nodes or broken links and located through a QW search.\\
Unfortunately the QW presented in the core of the paper is not robust to the presence of broken links, as it stands. To understand where the problem lies, comparison with the simpler 2 spatial dimensions scenario of Fig.~\ref{fig:2DDecoupling} helps. Ignore the grey arrows and suppose that one of the nodes is missing. This leaves two black arrows dangling, with nowhere to go. The amplitudes that they carry cannot be dropped as this would break unitarity. Nor can they be placed back in their originating nodes, as new amplitudes have come to occupy the space.\\
The only possible fix is to double up the number of amplitudes, and balance the flow of information by considering the grey arrows. It seems unlikely that we can achieve robustness without doubling up the number of amplitudes. In the absence of a node, we shall be able to plug back the dangling outcoming black arrow of a node into its dangling incoming grey arrow. In this appendix, we do the Tetrahedral QW counterpart of this construction, in order to reach a less optimised, but more robust scheme.

\subsection{Dynamics}

\begin{figure}
    \centering
    \includegraphics[width=1\columnwidth]{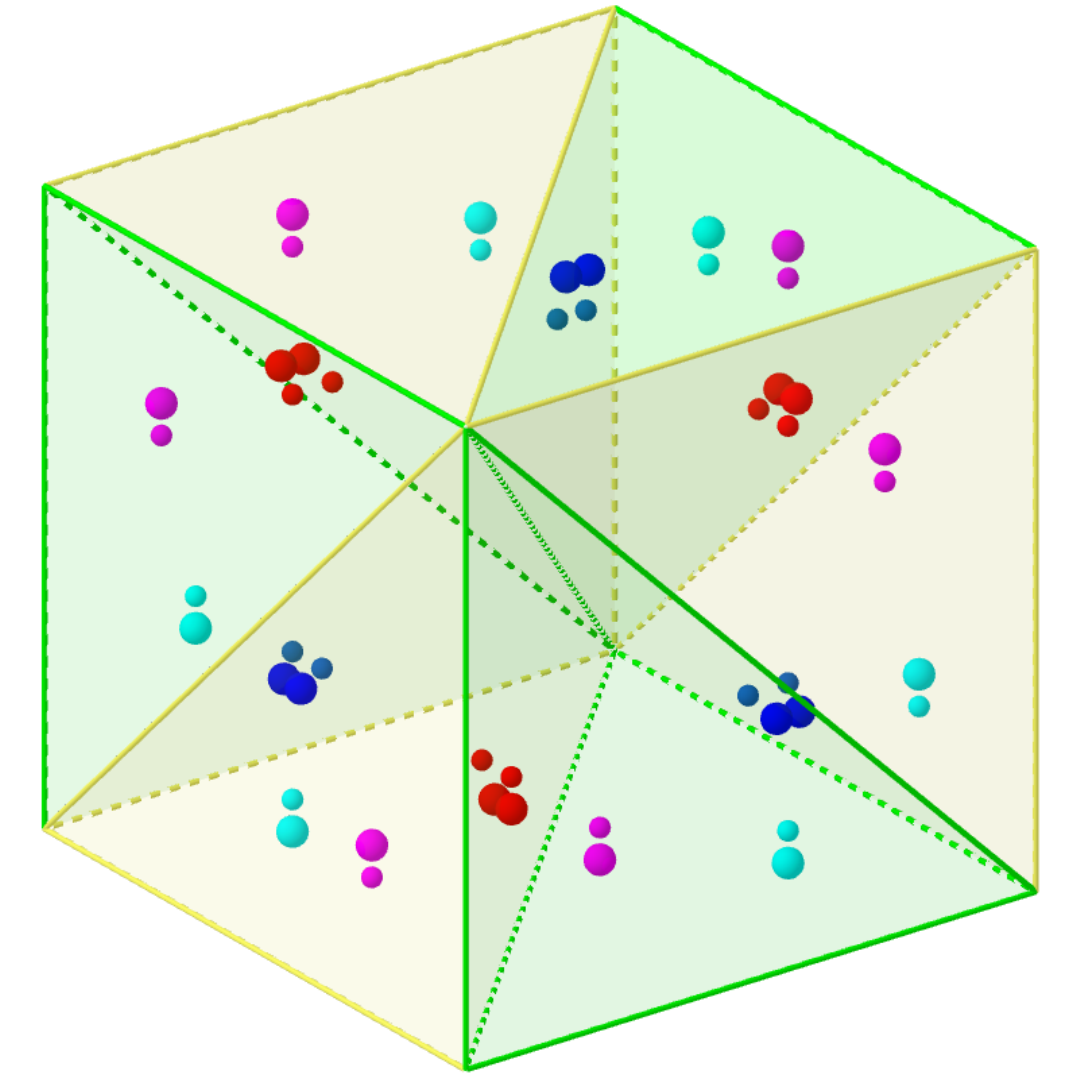}
    \caption{Dissection of a cube into six tetrahedra for the robust QW. The big (resp. small) dots are that of the first (resp. mirror) walker.}
    \label{fig:robust_cube}
\end{figure}

As explained above, we can define a robust QW by doubling the
number of amplitudes as shown in Fig. \ref{fig:robust_cube}, and
modifying the dynamics. We now only use swap operations for the shifting
dynamics. This robust QW contains two independent walkers. The mirror walker does the same displacements as the first walker but in the opposite direction. Each facet of a tetrahedron owns one amplitude of each walker. Recall that the two-dimensional spinors lie in the facet shared by neighboring tetrahedra and contain an up and down
component. Hence, each facet of a tetrahedron owns an up and down component
of the two different walkers. The state of tetrahedron $k$ at time
$t$ reads: 
\begin{equation}
    \phi(t,k) = (\phi(t,k,i,\uparrow),\phi(t,k,i,\downarrow))_{i=0,\dots,3}^T \quad .
\end{equation}
The blue, red, cyan, magenta dots of Fig. \ref{fig:robust_cube} correspond to $i=0,1,2,3$. In the shifting dynamics of the non-robust first walker, the amplitudes located on the LH (resp. RH) tetrahedra move in a positive (resp. negative) direction over time. Moreover, we know from Eq. \eqref{eq:movement2} that the up (resp. down) component of the non-robust first walker goes in negative (resp. positive) direction over time. Therefore, in the LH (resp. RH) tetrahedra, the up (resp. down) component is that of the mirror (resp. first) walker. Lastly, we define the $j$-th component of a tetrahedron as:
 \begin{equation}
     \phi_j(t,k)_{j=0,\dots,7} = 
     \begin{cases}
         \phi(t,k,j/2,\uparrow) \mbox{  if } j \mbox{ is even} \\
         \phi(t,k,(j-1)/2,\downarrow) \mbox{  otherwise}
     \end{cases}.
 \end{equation}

\subsubsection{Shifting operator}

\begin{figure}
    \centering
    \includegraphics[width=1\columnwidth]{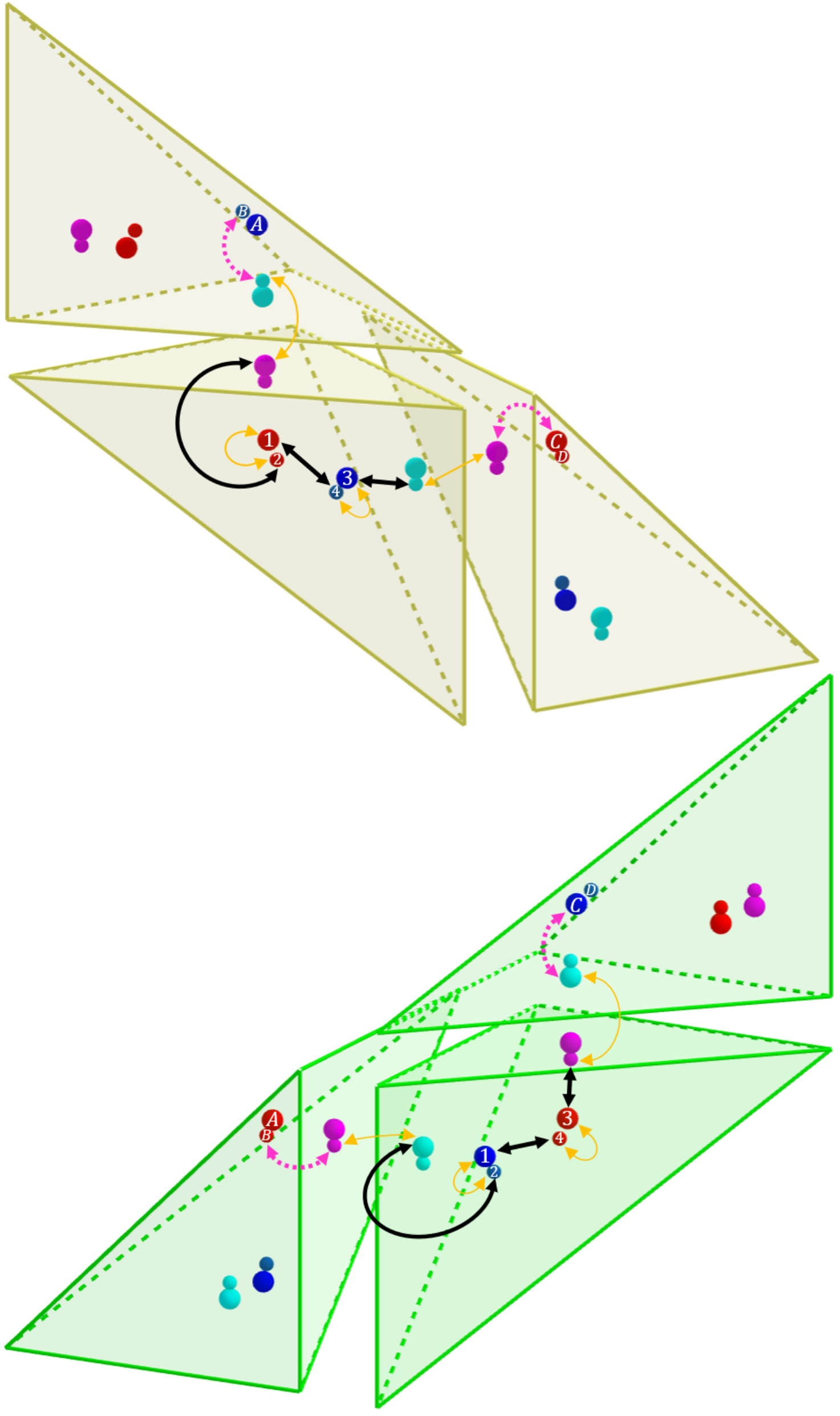}
    \caption{Shifting dynamics of the robust tetrahedral QW. The black (filled), orange (thin) and pink (dotted) arrows correspond to the first, second and third step needed to implement our shifting dynamics. We have labelled the amplitudes for readability, after the several displacements the amplitudes moved from: $1\rightarrow 3, 2\rightarrow B, 4\rightarrow 2$. Note that only the dynamics of the amplitudes of the middle tetrahedron is shown, thus, unrepresented displacements contain: $A\rightarrow 1, D\rightarrow 4$. Also, notice that only the coin step shuffles the amplitudes between LH and RH tetrahedra.}
    \label{fig:local_dynamics_robust}
\end{figure}

The shifting dynamics is done in 3 steps. Inspecting Fig. \ref{fig:local_dynamics_robust}, we see that for the LH (resp. RH) tetrahedra it consists in: (i) swap the mirror (resp. first) blue and the first (resp. mirror) cyan components, swap the first (resp. mirror) blue and mirror (resp. first) red components, swap the first (resp. mirror) red and mirror (resp. first) magenta components, (ii) swap the first and mirror blue components, swap the first and mirror red components, swap the first (resp. mirror) cyan and neighboring mirror (resp. first) magenta, swap the mirror (resp. first) magenta and the neighboring first (resp. mirror) cyan, (iii) swap the blue/cyan first (resp. mirror) components and swap the red/magenta mirror (resp. first) components.
The time evolution of a tetrahedron reads:
\begin{equation}
    \phi(t+\varepsilon,k) = (S_2S_1S_0\phi)(t,k) \quad ,
\end{equation}
where the $S_i$ operators formally act on a tetrahedron as:

\begin{subequations}
\begin{equation}
        (S_0\phi)(t,k) = 
    \begin{pmatrix}
        \phi_5(t,k) \\
        \phi_2(t,k) \\
        \phi_1(t,k) \\
        \phi_6(t,k) \\
        \phi_4(t,k) \\
        \phi_0(t,k) \\
        \phi_3(t,k) \\
        \phi_7(t,k)
    \end{pmatrix},
\end{equation}
\begin{equation}
     (S_1\phi)(t,k) = 
    \begin{pmatrix}
        \phi_1(t,k) \\
        \phi_0(t,k) \\
        \phi_3(t,k) \\
        \phi_2(t,k) \\
        \phi_4(t,k) \\
        \phi_6(t,n(k,2)) \\
        \phi_5(t,n(k,3)) \\
        \phi_7(t,k)
    \end{pmatrix},
\end{equation}
\begin{equation}
     (S_2\phi)(t,k) = 
    \begin{pmatrix}
        \phi_0(t,k) \\
        \phi_5(t,k) \\
        \phi_6(t,k) \\
        \phi_3(t,k) \\
        \phi_4(t,k) \\
        \phi_1(t,k) \\
        \phi_2(t,k) \\
        \phi_7(t,k)
    \end{pmatrix}.
\end{equation}
\end{subequations}

Let us compute the state of a tetrahedron after one time step:

\begin{equation}
\begin{split}(S_{2}S_{1}S_{0}\phi)(t,k) & =S_{2}S_{1}S_{0}\begin{pmatrix}\phi_{0}(t,k)\\
\phi_{1}(t,k)\\
\phi_{2}(t,k)\\
\phi_{3}(t,k)\\
\phi_{4}(t,k)\\
\phi_{5}(t,k)\\
\phi_{6}(t,k)\\
\phi_{7}(t,k)
\end{pmatrix}\\
=S_{2}S_{1} & \begin{pmatrix}\phi_{5}(t,k)\\
\phi_{2}(t,k)\\
\phi_{1}(t,k)\\
\phi_{6}(t,k)\\
\phi_{4}(t,k)\\
\phi_{0}(t,k)\\
\phi_{3}(t,k)\\
\phi_{7}(t,k)
\end{pmatrix}\\
=S_{2} & \begin{pmatrix}\phi_{2}(t,k)\\
\phi_{5}(t,k)\\
\phi_{6}(t,k)\\
\phi_{1}(t,k)\\
\phi_{4}(t,k)\\
\phi_{6}(t,n(k,2))\\
\phi_{5}(t,n(k,3))\\
\phi_{7}(t,k)
\end{pmatrix}\\
 & =\begin{pmatrix}\phi_{2}(t,k)\\
\phi_{6}(t,n(k,2))\\
\phi_{5}(t,n(k,3))\\
\phi_{1}(t,k)\\
\phi_{4}(t,k)\\
\phi_{5}(t,k)\\
\phi_{6}(t,k)\\
\phi_{7}(t,k)
\end{pmatrix}.
\end{split}
\end{equation}

We see that after one time step, only the blue $(j=0,1)$ and red $(j=2,3)$ components moved, while the others are left unchanged. Therefore, the global dynamics of our robust tetrahedral QW does not involve the cyan $(j=4,5)$ and magenta $(j=6,7)$ components. Thus, we only focus on the blue and red components. Although the cyan and magenta components do not move in the robust scheme after a time step, they are still necessary as ancillas in order to implement the shifting dynamics, as can be seen in Fig. \ref{fig:local_dynamics_robust}. They act as intermediaries between adjacent tetrahedra to avoid instantaneous displacements. However, after one time step, these components have not actually moved. Therefore, they are simply a tool contributing to the smooth running of the robust displacement dynamics. In the non-robust scheme, the cyan and magenta components were moving, which is not the case in the robust scheme. Restricting shifting operations to swaps makes it difficult to replicate the non-robust dynamics where all components move. Moreover, the robust dynamics that comprise two walkers moving in opposite directions is more intuitive and straightforward than the  walkers that follow each other of the non-robust scheme. Thus, it is more convenient to consider a simple dynamics involving two walkers moving in opposite directions, at the cost of immobilizing the cyan and magenta components, rather than trying to replicate the previous, less intuitive dynamic at all costs.

\begin{figure}
    \centering
    \includegraphics[width=1\columnwidth]{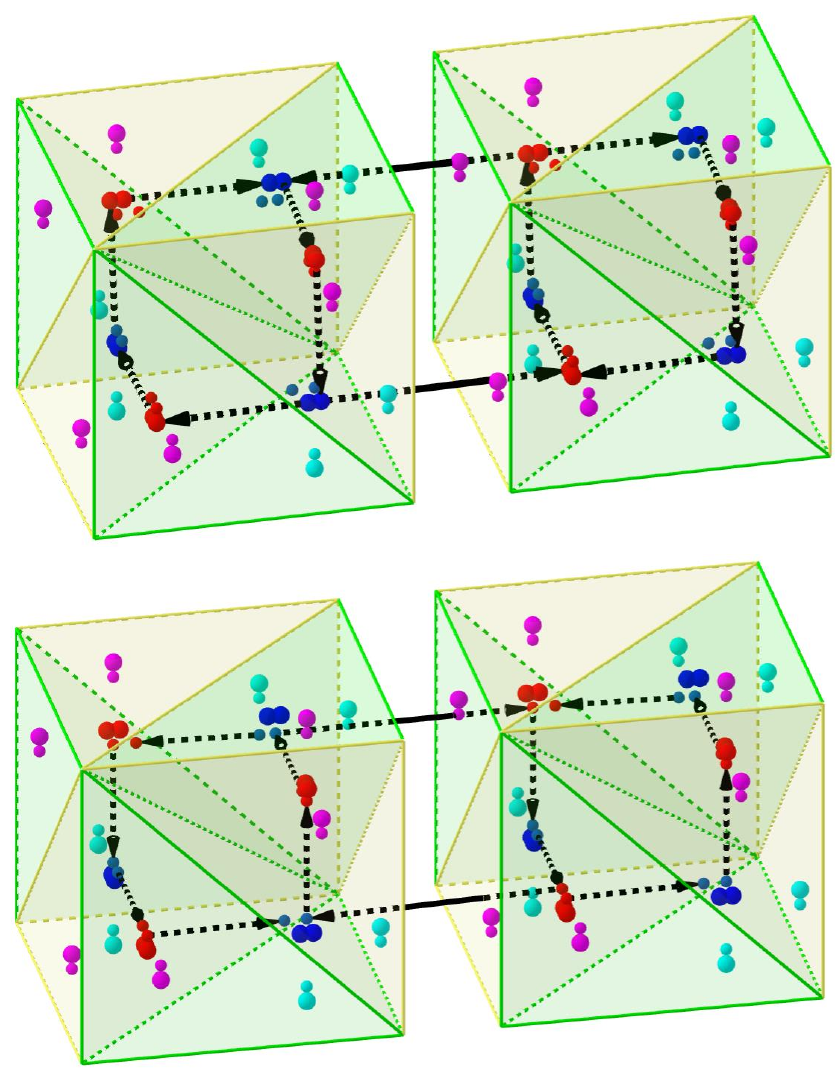}
    \caption{The upper (resp. lower) figure show the global dynamics of the first (resp. mirror) walker. The global shifting dynamics of the robust tetrahedral QW is just the fusion of the two figures.}
    \label{fig:global_dynamics_robust}
\end{figure}

\subsubsection{Coin operators}

The coin operator of the first walker is $C=e^{\mathrm{i}\frac{\pi}{3}}R_{\sigma_z}(\theta)R_{\sigma_x}(\theta)$ with $\theta=\pi/2$.

The mirror walker is doing the first walker walk in reverse. Thus, when applying the coin operator on the facets we want to map $\sigma_\mu$ into $\sigma_{\mu-1}$. Therefore, the coin of the mirror walker is $\tilde{C}=C^\dagger$.

In the following sections we will construct a robust QW that has the massive Dirac equation as its continuum limit. First, we will show
how a Weyl equation can be simulated in the continuum limit. By combining
two of such equations, we then can construct a robust QW for the massless
Dirac equation. Finally, we will introduce the mass term to prevent travelling at the speed of light.

\subsection{Weyl QW}

\subsubsection{First walker}

Each facet of a tetrahedron hosts a two dimensional spinor, let $\psi(t,k)$ (resp. $\tilde{\psi}(t,k)$) be that of the first (resp. mirror) walker. We formally define them as:
\begin{equation}
\psi(t,k)=\begin{cases}
\begin{pmatrix}\phi_{0}(t,n(k,0))\\
\phi_{1}(t,k)
\end{pmatrix}\mbox{ if }k\mbox{ is LH }\\
\begin{pmatrix}\phi_{2}(t,k)\\
\phi_{3}(t,n(k,1))
\end{pmatrix}\mbox{ otherwise }
\end{cases},
\end{equation}

\begin{equation}
\tilde{\psi}(t,k)=\begin{cases}
\begin{pmatrix}\phi_{0}(t,k)\\
\phi_{1}(t,n(k,0))
\end{pmatrix}\mbox{ if }k\mbox{ is LH }\\
\begin{pmatrix}\phi_{2}(t,n(k,1))\\
\phi_{3}(t,k)
\end{pmatrix}\mbox{ otherwise }
\end{cases}.
\end{equation}

Let us first derive the continuum limit of the first walker. Interestingly, the first walker was constructed to have the Weyl equation as its continuum limit. Recall the Weyl equation can be rewritten as: 
\begin{equation}\label{eq:ddiracfirst}
\begin{split}
    \mathrm{i}\partial_0 \psi &= D\psi\quad \textrm{with} \\
    D &= \mathrm{i}\sum_j\sigma_j\partial_j.
\end{split}
\end{equation}
It is known that a translation by $\varepsilon$ along the $\mu$-axis approximates as:
\begin{equation}
\tau_{\mu,\varepsilon}\psi = (\mathbb{I}+\varepsilon\partial_\mu)\psi + O(\varepsilon^2).
\label{eq:approxshift}
\end{equation}
Using the definition of $\sigma_3$ and Eq. (\ref{eq:shift}) we have:
\begin{equation}\label{eq:partial_shift_z}
    \left(  \mathbb{I}+\varepsilon \sigma_3\partial_3 \right) \psi = T_{3,\varepsilon}\psi+O(\varepsilon^2).
\end{equation}
Given that $\sigma_1 = C\sigma_3 C^\dagger$, we also have:
\begin{equation}
\left( \mathbb{I}+\varepsilon \sigma_1\partial_1 \right) \psi = C T_{1,\varepsilon} C^\dagger \psi+O(\varepsilon^2).
\end{equation}
Finally along the $y$-axis, as $\sigma_2 = C\sigma_1 C^\dagger=CC\sigma_3 C^\dagger C^\dagger$ and $C^2=C^\dagger$, we obtain:
\begin{equation}
\left( \mathbb{I}+\varepsilon \sigma_2\partial_2 \right) \psi = C^\dagger T_{2,\varepsilon} C\psi+O(\varepsilon^2).
\end{equation}
Lastly, the time evolution of the first walker is:
\begin{equation}\label{eq:firstwalk}
\psi(t+\varepsilon,k) = (W_\varepsilon\psi)(t,k)+O(\varepsilon^2),
\end{equation}
with:
\begin{equation}
     \begin{split}
         W_\varepsilon &= C^\dagger T_{2,\varepsilon}C CT_{1,\varepsilon}C^\dagger T_{3,\varepsilon} \\
         &= C^\dagger T_{2,\varepsilon}C^\dagger T_{1,\varepsilon}C^\dagger T_{3,\varepsilon}.
     \end{split}
\end{equation}
Using Eq. \eqref{eq:ddiracfirst} and Eq. \eqref{eq:approxshift} we have:
\begin{equation}\label{eq:psi2approx}
    \psi(t+\varepsilon,k) = (\mathbb{I}+\varepsilon \partial_0)\psi(t,k) + O(\varepsilon^2) .
\end{equation}
Moreover, with Eq. \eqref{eq:firstwalk} we get: 
\begin{equation}\label{eq:continuum_weyl}
    \begin{split}
        \psi(t+\varepsilon,k) &= \prod_{j=1}^3  C^\dagger (\mathbb{I}+\varepsilon\sigma_3\partial_j)\psi(t,k) + O(\varepsilon^2) \\
        \psi(t+\varepsilon,k) &= \prod_{j=1}^3 (\mathbb{I}+\varepsilon\sigma_j\partial_j)\psi(t,k) + O(\varepsilon^2) \\
        (\mathbb{I}+\varepsilon \partial_0)\psi(t,k) &= \left(\mathbb{I}+\varepsilon\sum_{j=1}^3\sigma_j\partial_j\right)\psi(t,k) + O(\varepsilon^2) \\
       \partial_0\psi(t,k) &= \sum_{j=1}^3\sigma_j\partial_j\psi(t,k) + O(\varepsilon^2).
    \end{split}
\end{equation}
In the continuum limit, we approximate the Weyl equation in (3+1)-dimensions, that is Eq.\ \eqref{eq:ddiracfirst}.

\subsubsection{Mirror walker}

We now focus on the mirror walker. The Weyl equation for the second chiral component has the form:
\begin{equation}
\begin{split}
    \mathrm{i}\partial_0 \tilde{\psi} &= \tilde{D}\tilde{\psi}\quad \textrm{with} \\
    \tilde{D} &= -\mathrm{i}\sum_j\sigma_j\partial_j.
\end{split}
\end{equation}
Doing the same computations as before, we get for the $z$-axis:
\begin{equation}
    \left( \mathbb{I}-\varepsilon \sigma_3\partial_3 \right) \tilde{\psi} = T_{3,\varepsilon}\tilde{\psi}+O(\varepsilon^2).
\end{equation}
Since $\sigma_1 = C\sigma_3 C^\dagger$, we also have:
\begin{equation}
\left( \mathbb{I}-\varepsilon \sigma_1\partial_1 \right) \tilde{\psi} = C T_{1,\varepsilon} C^\dagger\tilde{\psi}+O(\varepsilon^2).
\end{equation}
Lastly as $\sigma_2 = C^\dagger \sigma_3 C$, we obtain:
\begin{equation}
\left( \mathbb{I}-\varepsilon \sigma_2\partial_2 \right) \tilde{\psi} = C^\dagger T_{2,\varepsilon} C\tilde{\psi}+O(\varepsilon^2).
\end{equation}
The time evolution of the mirror walker is:
\begin{equation}\label{eq:mirrorwalker}
\tilde{\psi}(t+\varepsilon,k) = \tilde{W}_\varepsilon\tilde{\psi}(t,k),
\end{equation}
with:
\begin{equation}
     \begin{split}
         \tilde{W}_\varepsilon &=  C T_{1,\varepsilon} C^{\dagger} C^{\dagger}T_{2,\varepsilon} C T_{3,\varepsilon} \\
         &= C T_{1,\varepsilon} C T_{2,\varepsilon} C T_{3,\varepsilon}.
     \end{split}
\end{equation}

In the continuum limit we do obtain the seeked Weyl equation in (3+1)-dimensions since:
\begin{equation}
    \begin{split}
        \tilde{\psi}(t+\varepsilon,k) &= \prod_{j=1}^3 C(\mathbb{I}-\varepsilon\sigma_3\partial_j)\tilde{\psi}(t,k) + O(\varepsilon^2) \\
        \tilde{\psi}(t+\varepsilon,k) &= \prod_{j=1}^3 (\mathbb{I}-\varepsilon\sigma_j\partial_j)\tilde{\psi}(t,k) + O(\varepsilon^2) \\
        (\mathbb{I}+\varepsilon \partial_0)\tilde{\psi}(t,k) &= \left(\mathbb{I}-\varepsilon\sum_{j=1}^3\sigma_j\partial_j\right)\tilde{\psi}(t,k) + O(\varepsilon^2) \\
       \partial_0\tilde{\psi}(t,k) &= -\sum_{j=1}^3\sigma_j\partial_j\tilde{\psi}(t,k) + O(\varepsilon^2).
    \end{split}
\end{equation}

\subsection{Dirac QW}

\subsubsection{Massless Dirac QW}

The four-dimensional spinor $\Psi$ of the massless Dirac QW contains
all the amplitudes own by a facet where the first and second (resp.
third and fourth) components are that of the first (resp. mirror)
walker, that is $\Psi(t,k)=\psi(t,k)\oplus\tilde{\psi}(t,k)$. In
(3+1)-dimensions, the massless Dirac equation is: 
\begin{equation}
\begin{split}\mathrm{i}\partial_{0}\Psi & =\hat{D}\Psi\quad\textrm{with}\\
\hat{D} & =\sigma_{3}\otimes D\\
 & =\mathrm{i}\sum_{j=1}^{3}\alpha^{j}\partial_{j}
\end{split}
\quad,\label{eq:robustdirac}
\end{equation}
where $D$ is that of Eq. \eqref{eq:ddiracfirst} and $\alpha^j=\sigma_3\otimes\sigma_j$. The coin operator is $\hat{C}=\sigma_0 \otimes C$, as $\alpha^{j+1}=\hat{C}\alpha^j\hat{C}^\dagger$. Since the dynamics of our robust QW takes place in 3 steps, let $\mathcal{T}_{\mu,\varepsilon}^{(i)}$ be the $i$-th step of translation of $\varepsilon$ along the $\mu$-axis. We start with the $z$-axis:
\begin{equation}
    \left( \mathbb{I}+\varepsilon (\sigma_3\otimes-\sigma_3)\partial_3 \right) \Psi = \mathcal{T}_{3,\varepsilon}^{(2)}\mathcal{T}_{3,\varepsilon}^{(1)}\mathcal{T}_{3,\varepsilon}^{(0)}\Psi+O(\varepsilon^2).
\end{equation}
For the $y$-axis we get:
\begin{equation}
    \left( \mathbb{I}+\varepsilon (\sigma_3\otimes-\sigma_2)\partial_2 \right) \Psi = \hat{C}^\dagger\mathcal{T}_{2,\varepsilon}^{(2)}\mathcal{T}_{2,\varepsilon}^{(1)}\mathcal{T}_{2,\varepsilon}^{(0)}\hat{C} \Psi+O(\varepsilon^2).
\end{equation}
Lastly, for the $x$-axis we obtain:
\begin{equation}
    \left( \mathbb{I}+\varepsilon (\sigma_3\otimes-\sigma_1)\partial_1 \right) \Psi = \hat{C}\mathcal{T}_{1,\varepsilon}^{(2)}\mathcal{T}_{1,\varepsilon}^{(1)}\mathcal{T}_{1,\varepsilon}^{(0)}\hat{C}^\dagger\Psi+O(\varepsilon^2).
\end{equation}
The time evolution of the massless robust Dirac QW reads:
\begin{equation}
\Psi(t+\varepsilon,k) = (\hat{W}_\varepsilon\Psi)(t,k),
\end{equation}
with:
\begin{equation}
     \hat{W}_\varepsilon = \prod_{j=1}^3 \hat{C}\mathcal{T}_{j,\varepsilon}^{(2)}\mathcal{T}_{j,\varepsilon}^{(1)}\mathcal{T}_{j,\varepsilon}^{(0)}.
\end{equation}
We now verify that we approximate the Dirac equation:
\begin{equation}
    \begin{split} 
        \Psi(t+\varepsilon,k) &= \prod_{j=1}^3 \hat{C} (\mathbb{I}+\varepsilon(\sigma_3\otimes -\sigma_3)\partial_j)\Psi(t,k) + O(\varepsilon^2) \\
        \Psi(t+\varepsilon,k) &= \prod_{j=1}^3 (\mathbb{I}+\varepsilon(\sigma_3\otimes -\sigma_j)\partial_j)
        \Psi(t,k) + O(\varepsilon^2) \\
        (\mathbb{I}+\varepsilon\partial_0)\Psi(t,k) &= \left(\mathbb{I}+\sum_{j=1}^3\varepsilon(\sigma_3\otimes -\sigma_j)\partial_j\right)\Psi(t,k) + O(\varepsilon^2) \\
        \partial_0\Psi(t,k) &=
        \sum_{j=1}^3 (\sigma_3\otimes -\sigma_j)\partial_j\Psi(t,k)+ O(\varepsilon^2).
    \end{split}
\end{equation}
In the continuum limit we approximate Eq. \eqref{eq:robustdirac}.

\subsubsection{Massive Dirac QW}

We add the mass term to the Dirac QW in the same way we did in Sec.
\ref{subsec:diracmass}, namely by adding an additional mass coin.
Hence, the Dirac equation in Planck units reads: 
\begin{equation}
\begin{split}\mathrm{i}\partial_{0}\Psi & =\mathcal{D}\Psi\quad\textrm{with}\\
\mathcal{D} & =m\gamma^{0}+\mathrm{i}\sum_{j}\alpha^{j}\partial_{j}
\end{split}
\quad,\label{eq:robustmassdirac}
\end{equation}
where $m$ is the mass. The matrix $\gamma^{0}$ is that of Eq. \eqref{eq:alpha0}, i.e.
\begin{equation}
\gamma^{0}=\sigma_{1}\otimes\sigma_{0}.
\end{equation}
Let $M=e^{-\mathrm{i}\varepsilon m\gamma^{0}}$ be the associated mass coin. Moreover, we notice that
\begin{equation}
    \begin{split}
        \hat{C}M\hat{C}^\dagger &= M \\
        M\hat{C}M^{\dagger} &= \hat{C}
    \end{split},
\end{equation}
making it possible to apply both coins after translating along each axis. Hence, the time evolution of the robust Dirac tetrahedral QW is:
\begin{equation}
\Psi(t+\varepsilon,k) = (\mathcal{W}_\varepsilon\Psi)(t,k),
\end{equation}
where
\begin{equation}
    \label{eq:robustDiracQW}
     \hat{W}_\varepsilon = \prod_{j=1}^3 \hat{C}\mathcal{T}_{j,\varepsilon}^{(2)}\mathcal{T}_{j,\varepsilon}^{(1)}\mathcal{T}_{j,\varepsilon}^{(0)}M.
\end{equation}
Lastly, in the continuum limit we get: 
\begin{equation}
    \begin{split}
        \Psi(t+\varepsilon,k) &= (\mathbb{I}-\mathrm{i}\varepsilon \mathcal{D}) \Psi(t,k) + O(\varepsilon^2) \\
        &= \prod_{j=1}^3 \hat{C}^\dagger (\mathbb{I}+\varepsilon(\sigma_3\otimes \sigma_3)\partial_j)M\Psi(t,k) + O(\varepsilon^2) \\
        &= \prod_{j=1}^3 (\mathbb{I}+\varepsilon(\sigma_3\otimes \sigma_j)\partial_j)M\Psi(t,k) +O(\varepsilon^2).
    \end{split}
\end{equation}
Moreover, as $M=\mathbb{I}-\mathrm{i}\varepsilon m\gamma^0+O(\varepsilon^2)$, we have: 
\begin{equation}
    \begin{split}
         (\mathbb{I}+\varepsilon \partial_0)\Psi(t,k) &= \left(\mathbb{I}-3\mathrm{i}m\varepsilon\gamma^0+\sum_{j=1}^3 \varepsilon(\sigma_3\otimes \sigma_j)\partial_j\right)\Psi(t,k) \\ &+O(\varepsilon^2) \\
        \partial_0\Psi(t,k) &= \left(-3\mathrm{i}m\gamma^0+\sum_{j=1}^3 (\sigma_3\otimes \sigma_j)\partial_j)\right)\Psi(t,k) \\
        &+O(\varepsilon^2),
    \end{split}
\end{equation}
which approximates Eq. \eqref{eq:robustmassdirac}.

Finally, the robust Dirac tetrahedral QW is summarized in Eq. \eqref{eq:robustDiracQW}. It consists of: (i) applying a mass coin in a strictly localized manner, (ii) making three translations, where the second one is strictly localized, and the first and last ones are weakly localized, (iii) applying a directional coin that is weakly localized to the tetrahedra. 
The different stages can not be merged due to their slightly different locality. Indeed, even though the directional coin $\hat{C}$ and the second translation step $\mathcal{T}_{j,\varepsilon}^{(1)}$ are strictly local, and the remaining translations $\mathcal{T}_{j,\varepsilon}^{(0,2)}$ with the mass coin $M$ are weakly local, the merger is impossible, as the directional coin must be applied after the translations, and these do not commute.

\bibliographystyle{plain}
\bibliography{tetraref}

\end{document}